\documentclass[letterpaper,12pt]{article}
\usepackage[utf8]{inputenc}
\usepackage{lipsum}
\usepackage[margin=0.7in]{geometry}
\usepackage{mathptmx}
\usepackage{xfrac}
\usepackage{graphicx}
\usepackage{float}
\usepackage{color,soul}
\usepackage{subcaption} 
\usepackage{wrapfig}
\usepackage{amsmath}
\usepackage{amssymb}
\usepackage{epstopdf}
\usepackage{setspace}
\usepackage{mathtools, nccmath}
\usepackage{fancyhdr}
\usepackage{fixmath}

\usepackage[
    style=numeric-comp,
    bibstyle=phys,
   articletitle=true,
    biblabel=brackets,
    backend=biber,eprint=true,sorting=none
]{biblatex}
\usepackage{hyperref}
\hypersetup{
  colorlinks = true,
            allcolors  = red,
            citecolor  = blue}
\usepackage[compact]{titlesec}
    \titlespacing{\section}{0pt}{1ex}{1ex}
    \titlespacing{\subsection}{0pt}{1ex}{0ex}
    \titlespacing{\subsubsection}{0pt}{0.5ex}{0ex}
\usepackage{soul}
\DeclareUnicodeCharacter{0327}{,}

\def\bea{\begin{eqnarray}}
\def\eea{\end{eqnarray}}
\def\be{\begin{equation}}
\def\ee{\end{equation}}

\def\p{\partial}

\title{Effect of Gravitational Waves on Yang-Mills condensates}
\author{Narasimha Reddy Gosala \footnote{narasimha.gosala@uleth.ca, Corresponding Author}, Arundhati Dasgupta \footnote{arundhati.dasgupta@uleth.ca}\\
 4401 University Drive, Department of Physics and Astronomy,\\ University of Lethbridge, Canada.}
\date{\vspace{-5ex}}

\begin{document}

\maketitle

\begin{abstract}

In this article, we investigate the interactions of a Yang-Mills (YM) wave fluctuation of a classical isotropic, homogeneous YM condensate, which models gluon plasma, with a Gravitational Wave (GW). We re-analyse the study of fluctuations of the gluon plasma using vector decomposition of the gauge field into scalar, longitudinal, and transverse modes. We find that there is an energy transfer between isotropic gluon condensate and plasmon modes, but the effect is delayed due to GW, and dependent on the initial conditions.  We also studied quarks in the background of YM condensate and GW. We find that the quarks break the isotropy of the condensate and the GW couples quarks of different flavours of the $SU(2)$ group. Thus, the GW induces flavour fluctuations, which has interesting implications for experimental observations and quark-gluon plasma physics.

\end{abstract}

\maketitle


\section{\label{sec:level1} Introduction}

Yang-Mills (YM) fields are known to play an important role in understanding the dynamics of varied phenomena in both early universe and relativistic heavy ion collisions.  The non-perturbative YM fields are used in the study of quark confinement and chiral symmetry-breaking phenomena in Quantum Chromodynamics (QCD) \cite{YMint,YMint1,YMint2}. The self-interacting non-abelian gauge fields help in understanding Quark-Gluon Plasma (QGP) \cite{qgpreview,QGP,QGPbook} which is postulated to have formed in the early universe and is being observed in heavy-ion collisions. The classical YM fields help in the study of the non-equilibrium matter formed at early times in heavy-ion collisions before the formation of QGP \cite{glasma,glasma1}. YM fields also play a significant part in the study of many aspects of cosmology like Dark Energy \cite{DE,DE1} and gauge field-driven inflation \cite{gaugeflation,gaugeflation1}. Also, they are expected to play an important role in understanding QCD vacuum energy \cite{QCDvacuum,QCDvacuum1,QCDvacuum2}.

QGP is a dense state of matter, primarily consisting of quarks and gluons in an unconfined state. Most of the studies relating to the study of QGP are based on its thermodynamic properties which use the Bag model \cite{bagmodel}. There are some studies related to QCD phase transition, which led to the formation of QGP, based on effective field theories like the NJL model \cite{NJL}. Some studies related to investigating QGP assume quarks and gluons propagate as waves owing to their unconfined state. Since there were many processes that generate gravitational waves (GWs) in the early universe \cite{GWgen,GWgen1,LIGO,saulson}, it is quite natural to expect an influence of GWs on QGP or in general cosmological processes. This motivates us to study the de-confined YM fields in the presence of GWs. Along these lines, we studied the interaction of GWs with progressive YM fields in different cosmological backgrounds in \cite{gnr}. In this paper, we extend that analysis for a homogeneous and isotropic YM condensate configuration along with fluctuations in a GW background.

The study of QGP using condensate and fluctuations were carried out by several authors \cite{Prokhorov,con,con2,con3,con4,conden1}. Using effective action analysis, it is found that the true vacuum state of YM theory can be described by a nonzero chromomagnetic gluon condensate \cite{effact}. Considering the YM ground state as a condensate and studying the fluctuations around this ground state will give a better understanding of QGP. Along these lines, Prokhorov, Pasechnik, and Vereshkov   \cite{Prokhorov} have studied the dynamics of fluctuations in the presence of $SU(2)$ YM condensate. In their work, they considered a spatially homogeneous and isotropic component of gauge field as a condensate and an inhomogeneous component as YM wave modes. It was shown that the interactions between large condensate and small YM modes set off a significant energy transfer from condensate to wave modes. They found that the longitudinal YM modes which are un-physical in degenerate YM theory become physical particles, acquiring frequency and dispersion, after quantization due to the interactions between condensate and wave modes. Later, the authors studied YM condensates \cite{con4} in the $SU(3)$ case which has 3 $SU(2)$ subgroups, and which give rise to interaction among 3 different overlapping $SU(2)$ condensates. We found this model of the YM condensate to be most useful to study the interactions with GW. This model of the plasma and the plasmons allowed us to extend the work of \cite{gnr} which studies the interaction of GW with YM progressive waves to the study of the interaction of a GW with a YM plasma-plasmon system. We label the condensate as the `plasma' and the other modes as the plasmons (or particles). 

There have been studies on the generation of GWs from QCD phase transition by considering the phase transition as a first-order phase transition \cite{GWgen} while some considered a different process from the relaxation of gluon condensate during QCD phase transition \cite{GWQCDPT}. In this paper, we analyse the condensate of \cite{Prokhorov} in the presence of a GW. Our aim was to study how a GW can generate a plasmon fluctuation on the YM plasma. We find that symmetric transverse modes identified in \cite{Prokhorov} are induced by the GW, but the longitudinal modes are not induced to first order in the fields. To identify the longitudinal and transverse modes of the propagating vector fluctuations in a transparent way, we consider an ansatz which is different than in \cite{Prokhorov}. In this new ansatz, the condensate is taken as a homogeneous and isotropic solution of the $SU(2)$ YM, as in \cite{Prokhorov}, and a singlet under $SU(2)$ transformations. However, the other vector modes are isolated as $(\Phi^a )$ longitudinal modes and transverse ones ($ \chi^a_{\sigma}$), where $\sigma=1,2$, and $a=1,2,3$ are the internal directions. We then analyze the system using similar methods as \cite{Prokhorov}. We find that there is an energy exchange between condensate and wave modes happens for longer times than compared with \cite{Prokhorov} and after some time, there is an energy transfer from condensate to fluctuations as predicted in \cite{Prokhorov}. Note in most of our calculations, we use exact solutions to the system and do not use linearized approximations as in \cite{Prokhorov}.



Then, we studied the condensate + fluctuations system in the presence of GWs. We find that the GW can perturb the condensate and induce the `plasmon modes' or `particles' which carry away the energy. We also analyzed the system perturbatively and numerical solutions were obtained. The GW interacts with the longitudinal modes as well as the transverse modes. In the presence of GW, the plasmon oscillations were generated even when they were zero initially. This can be understood as the GW initiating condensate decay into plasmons. These GW wave-induced `plasmons' are obtained using both the longitudinal and transverse modes of the YM gauge fields. 

Since QGP has quarks, we studied the combined system of condensate and massless Fermions. There were many studies on the Fermions in the background of gauge fields during different scenarios \cite{fer,fer1,fer2,fer3,fer4}. The previous works in this area have focused on massless Fermions \cite{fer} and also on the production and back reaction of massive Fermions in spontaneously broken gauge theory \cite{fer1,fer2} and during axion inflation \cite{fer3,fer4}. In this work, we consider the massless Fermions in the background of an $SU(2)$ YM condensate. We find that Fermions break the isotropy assumption of the condensate upon back-reaction. 

Further, we studied the condensate + Fermions system in the presence of GWs. We find that the GW coupled to the quark-condensate system induces flavour fluctuations. This is an interesting aspect of the calculations, which will have consequences for the quantum theory of QGP. We are working to better understand the system \cite{condensate1}. In previous work \cite{grav}, there has been an attempt to attribute the low viscosity of the QGP to gravitons. Our calculations suggest that the GW background might shed light on this when one computes the QGP thermodynamics. In \cite{fermionscatter}, the graviton-induced flavour transitions have been studied and form factors obtained. In our calculations, the gravitational interaction is a classical wave, which is appropriate for the energies of the QGP production in RHIC. However, we expect new results when we use the new flavour-inducing transitions in the strangeness enhancement calculations of \cite{strange1,form}. QGP phase in heavy-ion collisions is known to lead to strange flavour heavy quark generation. Our GW interaction terms will add to the form factors of strangeness generation \cite{strange1,form}.

The aim of this paper is the study of the QGP behavior using a Yang-Mills condensate model and its fluctuations. The condensate decays into plasmons as a result of interactions. This paper uses a vector decomposition of the plasmons, and the decay is into longitudinal and transverse modes. The condensate by itself does not see the GW, but through the longitudinal and transverse modes there is an interaction with the GW. We also show that the gravitational wave can delay the decay of the condensate. This is because the plasmons that carry away the condensate energy do so at a lower rate, because a fraction of their energy gets used up in the interaction with the GW. Next, we show that adding quarks to the model breaks the isotropy of the condensate, and therefore a realistic QGP will have a non-isotropic condensate. A finite-temperature analysis of these observations of the Yang-Mills and quark system will provide more concrete predictions for the plasma, which is in progress \cite{condensate1}. We also show that the GW can induce flavour transitions in the quarks, in the background of the condensate. This is an important result and can be used to find new features of QGP in the early universe as well as in colliders, where weak GW might be present.

The paper is organized as follows: in Sec. \ref{sec:2}, we describe the condensate model of \cite{Prokhorov}, their linearized analysis, and show how a GW interacts with it. In Sec. \ref{sec:3}, we re-analyze the condensate system using a varied decomposition and use non-perturbative analysis. In Sec. \ref{sec:4}, we add Fermions to the system and describe how the quarks interact with the condensate and the GW. Finally, we conclude in Sec. \ref{sec:5}. In this paper, we work in the natural units: $\hslash =1,\ c=1$ and the Minkowski metric convention as $\eta_{\mu\nu} = {\rm diag}(-1,1,1,1)$.

\section{Fluctuations around Condensate in Linear-order approximation} {\label{sec:2}}
Consider the Yang-Mills (YM) theory with  a gauge invariant Lagrangian \cite{Peskin}
\begin{equation}
    \mathcal{L} = - \frac{1}{4} F^{a\mu\nu} F^a_{\mu\nu},
\end{equation}
where $F^a_{\mu\nu}$ is the anti-symmetric field strength tensor of gauge field $A^a_\mu $ given by 
\begin{equation}{\label{eqF}}
    F^a_{\mu\nu} = \partial_\mu A^a_\nu - \partial_\nu A^a_\mu + g_{ym} f^{abc} A^b_\mu A^c_\nu,
\end{equation}
where $g_{ym}$ is the YM coupling constant and $ F_{\mu\nu} = F^{a}_{\mu\nu} T^a, $ with the generators $T^a$ of Lie algebra $\mathfrak{su}(N)$, indexed by $a$, satisfying $ {\rm Tr}(T^a T^b) = \frac{1}{2} \delta^{ab}, [T^a,T^b] = i f^{abc} T^c $ here $f^{abc}$ are the structure constants of Lie algebra. Greek indices represent the space-time coordinates. Among Latin indices, $a,b,c,...$ represent the internal indices of Lie algebra $\mathfrak{su}(N)$ and $i,j,k,l,...$ represents the spatial coordinates. In our work, we are working with the $SU(2)$ Lie group. In this case, the Latin indices $a,b,c,..$ restricts to $1,2,3$ only, and the structure constants become Levi-Civitia tensor, i.e, $f^{abc} = \epsilon^{abc}$. It means the Latin indices $a,b,c,..$ and $i,j,k,..$ take the same range of values. The generators of the $\mathfrak{su}(2)$  algebra are $T^a=\sigma^a/2\ (\sigma^a = {\rm Pauli\ matrices})$.

\begin{figure}[htbp]
    \begin{subfigure}{0.4\textwidth}
      \includegraphics[width=\textwidth]{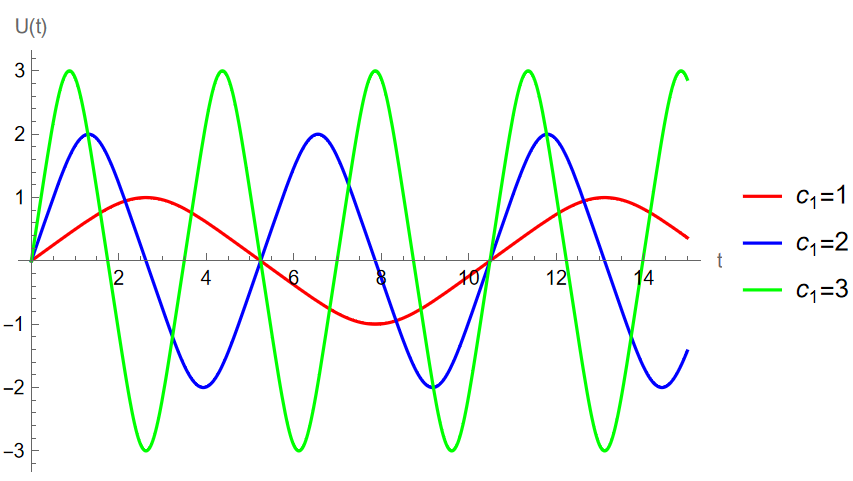}
      \caption{$U(t)$ at $c_2=0$}
      \label{fig:unUc1}
    \end{subfigure}
    \hfill
    \begin{subfigure}{0.4\textwidth}
      \includegraphics[width=\textwidth]{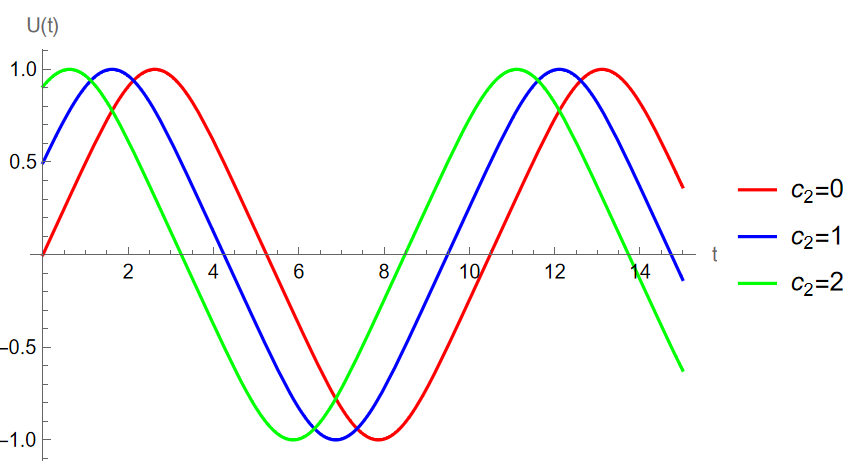}
      \caption{$U(t)$ at $c_1=1$}
      \label{fig:unUc2}
    \end{subfigure}
    \hfill
    \caption{Figures showing $U(t)$ for different $c_1$ and $c_2$ values. We choose $g_{ym}=0.5$. }
    \label{fig:Udiffc1c2}
\end{figure}

The equations of motion corresponding to the above Lagrangian are 
\begin{equation}{\label{eqYM}}
    \nabla_\mu F^{a \mu\nu} + g_{ym} \  \epsilon^{abc}\  A_{\mu}^{b}\  F^{c\mu\nu} = 0,
\end{equation}
where $\nabla_{\mu}$ is the covariant derivative. As suggested in \cite{Dyadichev, Galtsov, Galtsov1991, Cembranos}, the  SU(2) YM field can be represented in terms of a homogeneous isotropic configuration parametrized by a single scalar function because the SU(2) group is locally isomorphic to the SO(3) group of spatial rotations and their Lie algebras are same. From here on, we consider the Latin indices for both internal indices of SU(2) and for space indices. As in the work of \cite{Prokhorov}, we considered the following ansatz in the Hamilton gauge, $A^a_0=0$,
\begin{equation}
    A^a_i=U(t)\delta^a_i+\delta^{aj} \tilde{A}_{ji}(t,\vec{x}),
    \label{eqn:fluct1}
\end{equation}
where  $U(t)$ is the spatially homogeneous time-dependent scalar field corresponding to the YM condensate and $\tilde{A}_{ik}$ is the spatially in-homogeneous fluctuations components. When the fluctuations are set to zero, the YM equation (\ref{eqYM}) reduces to 
\begin{equation}
    \partial_0 \partial_0 U +2 g_{ym}^2 U^3 = 0.
\end{equation}
The exact solution for the above equation is given in terms of Jacobi elliptic functions as
\begin{equation}{\label{eq:zero}}
    U(t) = c_1 \ {\rm sn}(g_{ym} c_1 (t +c_2),-1), 
\end{equation}
where ${\rm sn}(g_{ym} c_1 (t +c_2),-1)$ is the sine Jacobi elliptic function \footnote{Here, the Sine Jacobi elliptic function ($y(u) =\rm{sn}(u,k^2)$) is defined as a solution of non-linear differential equation: $ \frac{d^2 y}{du^2} + (1+k^2) y- 2 k^2 y^3 =0$. It is also defined in terms of the inverses of elliptic functions as follows: $\rm{sn}(u,k^2)$ = $\sin \varphi$ where $\varphi$ satisfies the following equation: $ u = \int^\varphi_0 \frac{d\theta}{\sqrt{1-k^2 \sin^2\theta}}$} and $c_1$ and $c_2$ are the constants of integration (We used the notation $\rm{sn}(u,k^2)$, where $k$ is the elliptic modulus). Out of two constants, one regulates the energy of the condensate and the other regulates the phase of the condensate. This is given in Fig. (\ref{fig:Udiffc1c2}). 

Now, consider the linear order approximation in which only the interaction between condensate and fluctuations was considered. The YM equations were given by 
\begin{equation}
    \Box \tilde{A}_{ij} -\partial_k \partial_j \tilde A_{ik}  +  g_{ym} U \big(2\epsilon_{ikc} \partial_k \tilde{A}_{cj}  + \epsilon_{ibj} \partial_k \tilde{A}_{bk}  + \epsilon_{ibc} \partial_j \tilde{A}_{bc}\big) + g_{ym}^2 U^2 \left(\tilde{A}_{ji} - \tilde{A}_{ij}-2 \tilde{A}_{kk} \delta_{ij} \right) =0 .   
\end{equation}
where the d'Alembertian operator is defined as $\Box = \partial_\mu \partial^\mu$. The Gauss law obtained from the $\nu=0$ component in the YM equation (\ref{eqYM}), acts as a constraint. In the linearized form, it is given by 
\begin{equation}
    \partial_k \partial_0 \tilde{A}_{ik}+g_{ym} \partial_0 U \epsilon_{ijk}\tilde{A}_{jk} +g_{ym} U \epsilon_{ikj}\partial_0 \tilde{A}_{jk} = 0.
\end{equation}
In \cite{Prokhorov}, the fluctuations $\tilde{A}_{ij}$ are decomposed into longitudinal and transverse modes.  In the following, we analyze the symmetric transverse trace-less modes only, as the GW interacts with only these. The GW does not couple to the longitudinal modes in this particular analysis \cite{Prokhorov}.
The trace-less modes of the fluctuations are
\begin{equation}
    \tilde{A}_{ij}= Q^\alpha_{ij} \Psi_\alpha, 
\end{equation}
where the coefficient $Q^\alpha_{ij}$ satisfies the following conditions: $Q^\alpha_{ij}= Q^\alpha_{ji}$, $Q^\alpha_{ii}=0$ and $\alpha$ take the values 1 and 2. If the YM fluctuation mode is propagating in the $z$-direction, then the only non-zero components of $Q^\alpha_{ij}$ are $Q^1_{11}=-Q^1_{22}=1$ and 
$Q^2_{12}=Q^2_{21}=1$. With this choice of modes, the constraint equation gets automatically satisfied and the equations for the modes $\Psi_\alpha$ are given by 
\begin{align}{\label{eq:psinogw}}
     \Box \Psi_1 - 2 g_{ym} U \partial_z \Psi_2&=0, \\ 
     \label{eq:psinogw1}
     \Box \Psi_2 + 2 g_{ym} U \partial_z \Psi_1&=0. 
\end{align}
In Fourier space in the spatial coordinates ($e^{i p z}$), these equations can be written as generalized Mathieu equations. These equations were well studied in the case of parametric resonance \cite{hill,ivana,ivana1}.

\begin{figure}[htbp]
\begin{subfigure}{0.3\textwidth}
      \includegraphics[width=\textwidth]{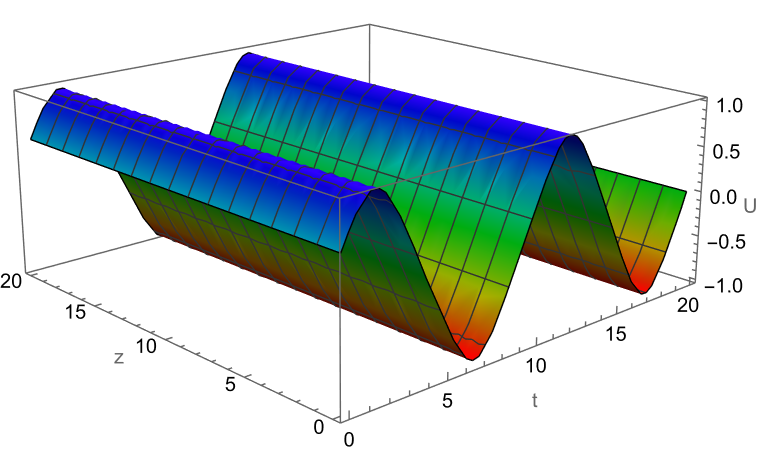}
      \caption{$U(t)$ for every $z$}
      \label{fig:unU}
    \end{subfigure}
    \hfill
    \begin{subfigure}{0.3\textwidth}
      \includegraphics[width=\textwidth]{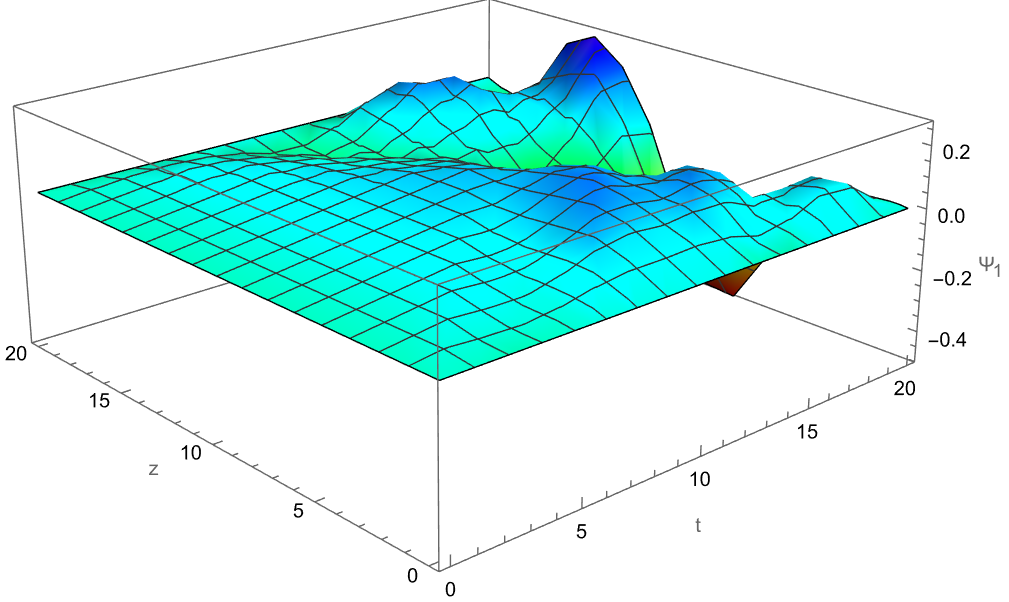}
      \caption{$\Psi_1(t,z)$}
      \label{fig:psi1}
    \end{subfigure}
    \hfill
    \begin{subfigure}{0.3\textwidth}
      \includegraphics[width=\textwidth]{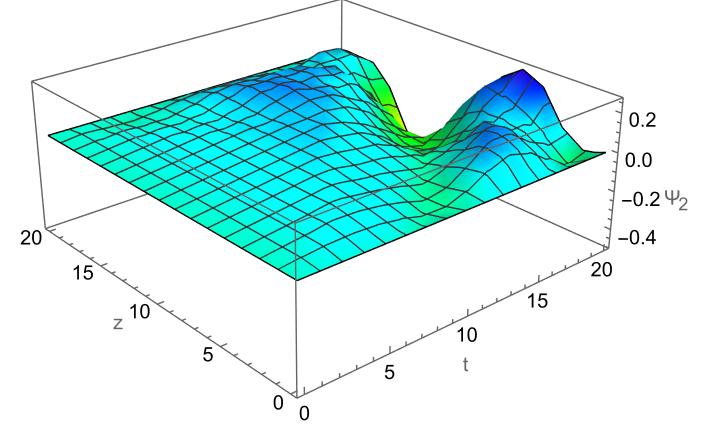}
      \caption{$\Psi_2(t,z)$}
      \label{fig:psi2}
    \end{subfigure}
    \hfill
    \caption{Figures showing condensate and transverse-traceless modes in the absence of gravitational wave. Fig. (a) represents $U(t)$ at every $z$. Figs. (b) \& (c) represents $\Psi_1(t,z)$ and $\Psi_2(t,z)$ in the presence of condensate. We choose $g_{ym}=0.5$, $c_1=c_2=1$, $\Psi_1(0,z)=\Psi_1(t,0) = \Psi_1(t,20) = 0.001$ and $\Psi_2(0,z)=\Psi_2(t,0) = \Psi_2(t,20) = 0.001$.}
    \label{fig:psi}
\end{figure}

We solved the above equations using the Method of Lines, a numerical technique to solve PDEs (see Appendix  \ref{MOL}). The numerical solutions for $\Psi_\alpha$ are shown in Figs. (\ref{fig:psi1})  \& (\ref{fig:psi2}).

Without the Fourier transform as in \cite{Prokhorov}, we numerically obtained solutions for the transverse modes. Initially, both functions show almost similar behaviour. As time increases, the $\Psi_1(t,z)$ behaves differently than $\Psi_2(t,z)$. The question we ask is if GW can induce these modes from the condensate at $t=0$. The above would correspond to the system of fluctuations studied in \cite{Prokhorov}, with only the $\Psi_{\alpha}$, with non-zero initial boundary conditions. We study a simplified version of \cite{Prokhorov}, as the GW only couples or interacts with the symmetric transverse traceless tensor modes of $\tilde{A}_{ik}$. This is expected due to the nature of the GW which is a symmetric transverse traceless tensor.

\subsection{In the presence of Gravitational waves}
In this subsection, we want to study the fluctuations around a condensate in the presence of a GW. Here, we start with the YM equations in a general metric as in  \cite{gnr} 
\begin{equation}
    \frac{1}{\sqrt{-g}} \partial_{\mu} \left(\sqrt{-g} g^{\mu \lambda} g^{\nu \rho} F^{a}_{\lambda \rho}\right) + g_{ym} \epsilon^{abc} A_{\mu}^b F^{\mu \nu c}=0.
\end{equation}

In the linearized approximation, using the GW metric i.e. $g_{\mu\nu}=\eta_{\mu\nu}+h_{\mu\nu}$, we can split the equation in the following form, where we have taken the GW wave-dependent terms to one side as follows
\begin{multline}
    \eta^{\mu \lambda} \eta^{\nu \rho} \partial_{\mu} F^a_{\lambda \rho} + g_{ym} \epsilon^{abc} A_{\mu}^b F^{\mu \nu c}= g_{ym} \epsilon^{abc} A^b_{\mu} h^{\mu \lambda} \eta^{\nu \rho} F^c_{\lambda \rho}  + g_{ym} \epsilon^{abc} A_{\mu}^b h^{\nu \rho} \eta^{\mu \lambda} F^c_{\lambda \rho}  \\ +  \eta^{\nu \rho} \partial_{\mu} F^a_{\lambda \rho} h^{\mu \lambda} + \eta^{\mu \lambda} \partial_{\mu} (h^{\nu \rho} F^a_{\lambda \rho}). 
\end{multline}
We then solve the above equations using the same form of ansatz as in the previous section. In Hamilton gauge $A^a_0=0$, the ansatz is written as 
\begin{equation}{\label{eq:YMGW}}
        A_i^a= U(t) \delta^a_i +  \delta^{aj} \tilde{A'}_{ji},
\end{equation} 
where $\tilde{A'}_{ki}$ is the fluctuations component of the gauge field due to its interaction with condensate and gravitational wave which is different from $\tilde{A}_{ki}$ of Eq. (\ref{eqn:fluct1}), whose dependence was on the condensate only. In linear-order approximation, the YM equation can be written as 

\begin{multline}{\label{eq:YMGWEOM}}
    \Box \tilde{A'}_{ij} -\partial_k \partial_j \tilde A'_{ik}  +  g_{ym} U \big(2\epsilon_{ikc} \partial_k \tilde{A'}_{cj}  + \epsilon_{ibj} \partial_k \tilde{A'}_{bk}  + \epsilon_{ibc} \partial_j \tilde{A'}_{bc}\big) \\ + g_{ym}^2 U^2 \left(\tilde{A'}_{ji} - \tilde{A'}_{ij}-2 \tilde{A'}_{kk} \delta_{ij} \right) =   -\partial_0 (U h_{ji})+g_{ym} U^2 \epsilon_{ikl} \partial_k h_{jl}-g^2_{ym} U^3 h_{ji}.  
\end{multline}
We consider a plus-polarised GW to be propagating in the $z$-direction i.e. $h_{ij}=e^+_{ij} h_+$, where $e^+_{ij}$ is the polarisation vector of the gravitational wave with two non-zero values, $e^+_{11}=-e^+_{22}=1$  and $h_+(t,z)=A_+ \cos(\omega_g (t-z))$. Similarly, we consider only the same modes as in the previous section i.e. $\tilde{A'}_{ij} = Q^\alpha_{ij} \Psi'_\alpha$. Then, the YM equation reduces to 
\begin{eqnarray}{\label{eq:psiGW}}
     \Box \Psi'_1 - 2 g_{ym} U \partial_z \Psi'_2&=& A_+( g_{ym}^2 U^3 \cos(\omega_g(t-z))+ \partial_0 U \ A_+ \omega_g \sin(\omega_g(t-z)), \\     
     \label{eq:psiGW1}
     \Box \Psi'_2 + 2 g_{ym} U \partial_z \Psi'_1&=& -A_+ g_{ym} \omega_g U^2 \sin(\omega_g(t-z)). 
\end{eqnarray}

\begin{figure}[htbp]
    \begin{subfigure}{0.4\textwidth}
      \includegraphics[width=\textwidth]{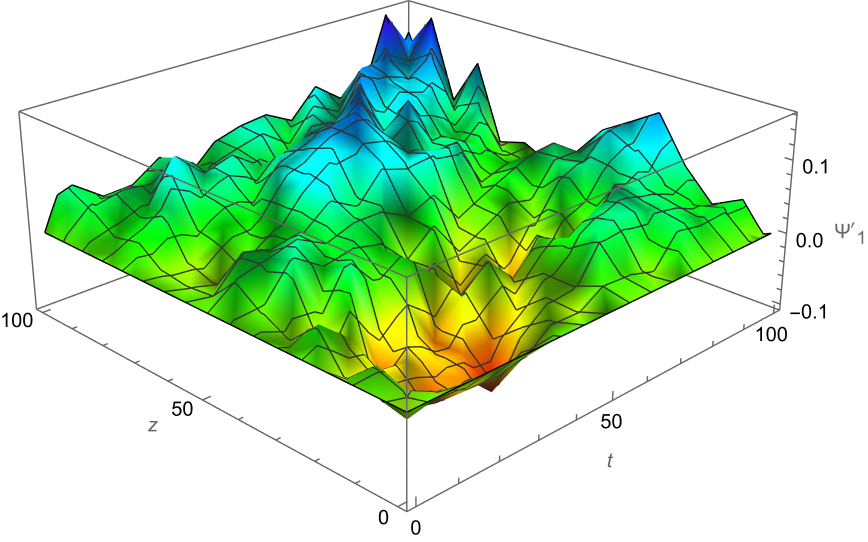}
      \caption{$\Psi'_1(t,z)$}
      \label{fig:psi1GW}
    \end{subfigure}
    \hfill
    \begin{subfigure}{0.4\textwidth}
      \includegraphics[width=\textwidth]{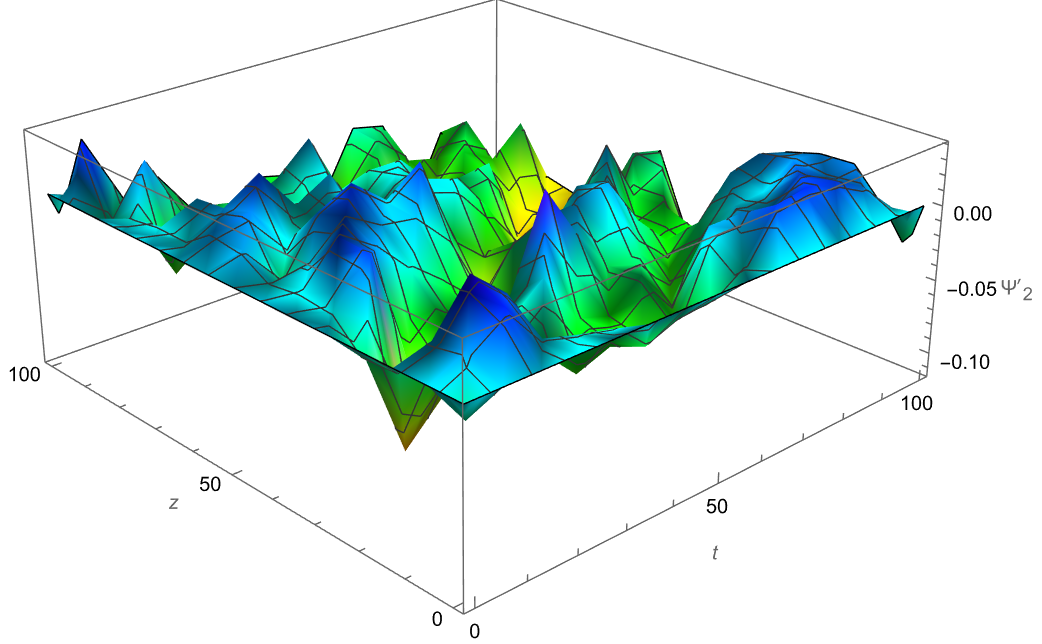}
      \caption{$\Psi'_2(t,z)$}
      \label{fig:psi2GW}
    \end{subfigure}
    \hfill
    \caption{Figures showing $\Psi'_1(t,z)$ and $\Psi'_2(t,z)$ in the presence of a gravitational wave. We choose $g_{ym}=0.5$, $c_1=c_2=1$, $A_+ = 0.0001$, $\omega_g=100$, $\Psi'_1(0,z)=\Psi'_1(t,0) = \Psi'_2(0,z)=\Psi'_2(t,0) = 0 $ and $ \partial_z \Psi'_1(t,0) = \partial_z \Psi'_2(t,0) = 0 $.}
    \label{fig:psiGW}
\end{figure}

We can solve the above equations and the numerical solutions are shown in Figs. (\ref{fig:psi1GW}) \& (\ref{fig:psi2GW}).    

Apart from the numerical solutions, we can also find the analytical solution by using the triad formalism. For this purpose, the gauge field can be rewritten using triads in the form of  given in \cite{Prokhorov},
\begin{equation}{\label{eq:YMtriads}}
        A_i^a= U(t) e^a_i + \ e^{aj}  \tilde{A}_{ji},
\end{equation} 
where $e_i^a$ are soldering forms that connect the space-time metric to the SU(2) group manifold which satisfies the following conditions:
\begin{equation}
        e^a_i e_{aj}= g_{ij};\hspace{1cm} e_i^a e^i_b= \delta^a_b.
\end{equation}
From the previous work \cite{Arundhati}, we take the triads in the case of a plus-polarised GW, to be 
\begin{subequations}
\label{eq:triads}
\begin{equation}
 e^a_1= \left( \sqrt{\left(1+ h_+(t,z)\right)},0,0\right),
\end{equation}
\begin{equation}
 e^a_2= \left(0, \sqrt{\left(1-h_+(t,z)\right)},0\right),
\end{equation}
\begin{equation}
 e^a_3=\left(0,0,1\right).
\end{equation}
\end{subequations}
Considering a binomial expansion of fractional powers and restricting only to first-order terms, the triads will be in the linear order of $h_+$. Using this and keeping the only terms to linear order in $h_+$ and $\tilde{A}^a_i$, one can write the gauge field (Eq. (\ref{eq:YMtriads})) as
\begin{equation}{\label{eq:YMGW1}}
    A^a_i=U(t) \delta^a_i  + \frac{1}{2} h^a_i U + \tilde{A}^a_{i}.
\end{equation}
If we compare the above equation with Eq. (\ref{eq:YMGW}), we find that the gauge field fluctuation in the presence of GW is related to the fluctuation without GW as
\begin{eqnarray}
     \tilde{A'}^a_i= \frac{1}{2} h^a_i U + \tilde{A}^a_i.
\end{eqnarray}
One can check that this solution indeed satisfies Eq. (\ref{eq:YMGWEOM}). In terms of $\Psi_\alpha$ and $\Psi'_\alpha$, we found that 
\begin{eqnarray}
    \Psi'_\alpha=\Psi_\alpha + \frac{1}{4} Q^\alpha_{ij} h_{ij} U.
\end{eqnarray}
In explicit forms, we get 
\begin{equation}
        \Psi'_1=\Psi_1 + \frac{1}{2} h_+ U ;\hspace{1cm} \Psi'_2=\Psi_2.
        \label{eq:fluc}
\end{equation}

In this simple analysis, we found an exact analytical solution for the fluctuation of the gauge field in the presence of both condensate and gravitational waves.
One can see that even if at $t=0$ the $\Psi_1,\Psi_2$ are zero, there is a $\Psi'_1$ induced due to the GW and the condensate interaction as in Eq. (\ref{eq:fluc}). 

\subsection{Higher order corrections}
\begin{figure}[ht]
    \begin{subfigure}{0.4\textwidth}
      \includegraphics[scale=0.4]{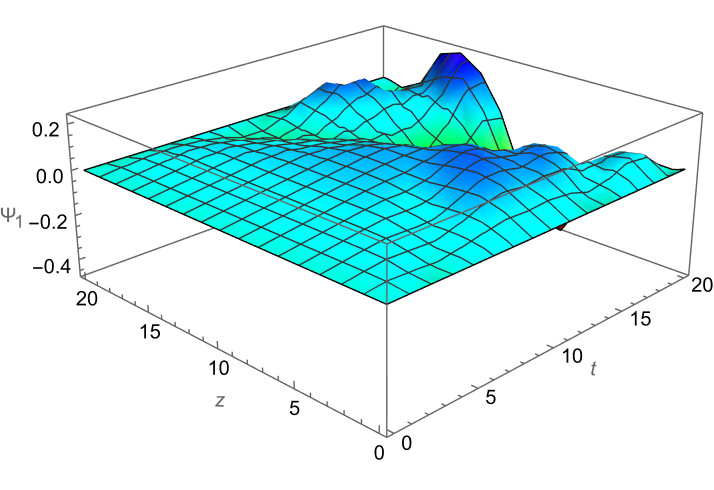}
      \caption{$\Psi_1(t,z)$}
      \label{fig:psi1all}
    \end{subfigure}
    \hfill
    \begin{subfigure}{0.4\textwidth}
      \includegraphics[scale=0.4]{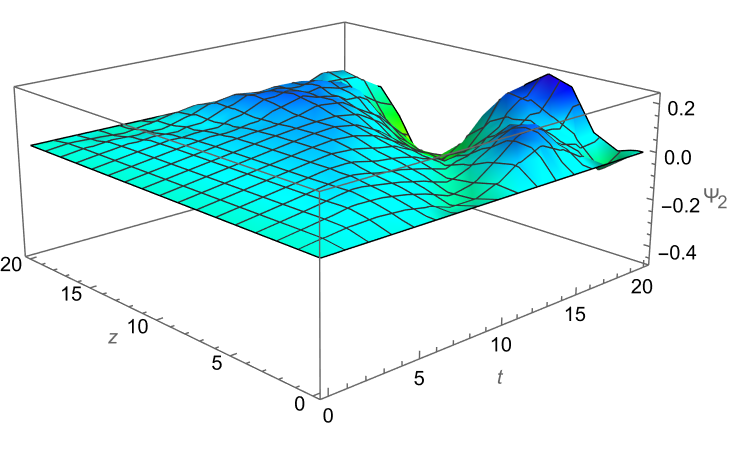}
      \caption{$\Psi_2(t,z)$}
      \label{fig:psi2all}
    \end{subfigure}
    \hfill
    \caption{Figures showing $\Psi_1(t,z)$ and $\Psi_2(t,z)$ in the presence of condensate to all orders in the self-interactions. We choose $g_{ym}=0.5$, $c_1=c_2=1$, $\Psi_1(0,z)=\Psi_1(t,0)=\Psi_1(t,20)=0.001$, $ \Psi_2(0,z)=\Psi_2(t,0)=\Psi_2(t,20)=0.001$.}  
    \label{fig:psiall}
\end{figure}

\begin{figure}[ht]
    \begin{subfigure}{0.4\textwidth}
      \includegraphics[width=8cm,height=4cm]{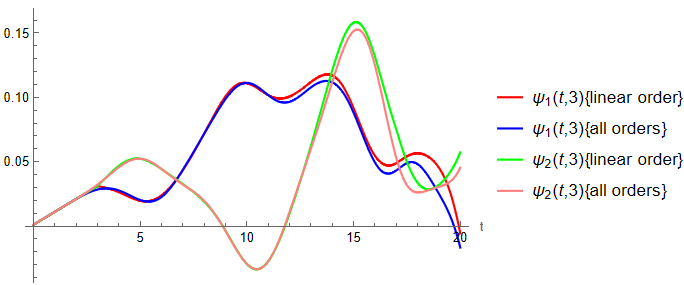}
      \caption*{}
    \end{subfigure}
    \hfill
    \begin{subfigure}{0.4\textwidth}
      \includegraphics[width=8cm,height=4cm]{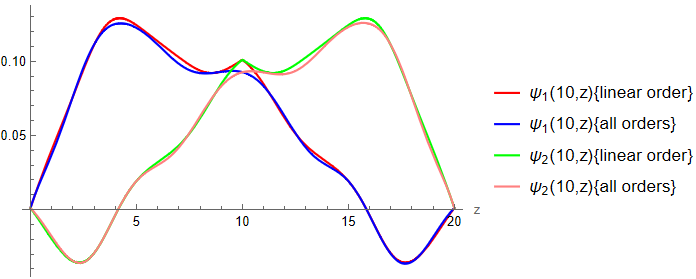}
      \caption*{}
    \end{subfigure}
    \caption{Figures showing $\Psi_1(t,z)$ and $\Psi_2(t,z)$ in the presence of condensate as functions of $t$ and $z$ separately. The red and green plots corresponds to linear order in interactions, while the blue and pink plots corresponds to all orders in self-interactions. We choose $g_{ym}=0.5$, $c_1=c_2=1$, $\Psi_1(0,z)=\Psi_1(t,0)=\Psi_1(t,20)=0.001$, $ \Psi_2(0,z)=\Psi_2(t,0)=\Psi_2(t,20)=0.001$.}  
    \label{fig:psipsiallcomp}
\end{figure}

If one wants to work beyond linear order approximation as in \cite{Prokhorov}, then one has to consider the interactions between the modes. Taking into account higher-order terms, the equations for $\Psi_1$ and $\Psi_2$ (Eqs. \ref{eq:psinogw} \& \ref{eq:psinogw1}) changed to 
\begin{align}{\label{eq:psinogwall}}
     \Box \Psi_1 - 2 g_{ym} U \partial_z \Psi_2-g_{ym}^2 (\Psi_1^3+\Psi_1 \Psi_2^2)&=0, \\ 
     \Box \Psi_2 + 2 g_{ym} U \partial_z \Psi_1-g_{ym}^2(\Psi_2^3+\Psi_1^2 \Psi_2)&=0.
\end{align}
The solutions for the above equations are given in Fig. (\ref{fig:psiall}). As the change due to non-linear terms is very small, we plotted 2D plots to show the effect of non-linear terms in Fig. (\ref{fig:psipsiallcomp}).

\begin{figure}[htbp]
    \begin{subfigure}{0.4\textwidth}
      \includegraphics[scale=0.4]{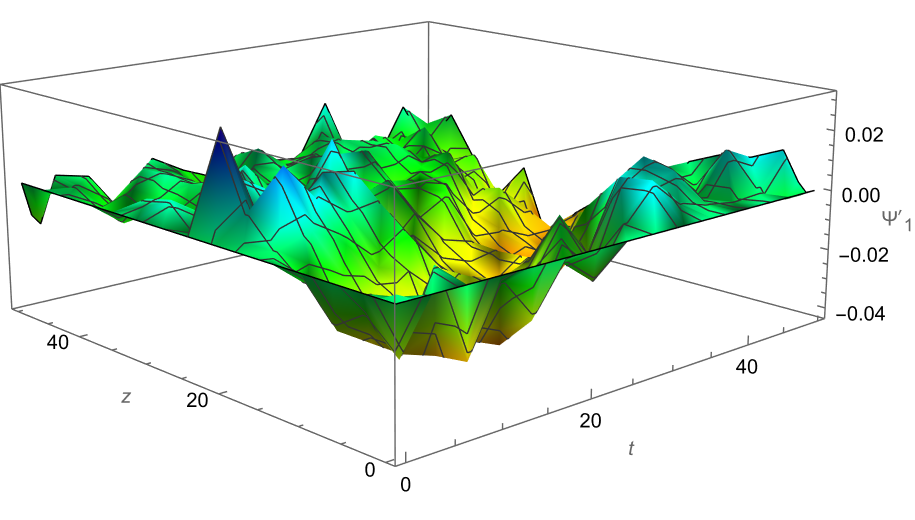}
      \caption{$\Psi'_1(t,z)$}
      \label{fig:psi1allGW}
    \end{subfigure}
    \hfill
    \begin{subfigure}{0.4\textwidth}
      \includegraphics[scale=0.4]{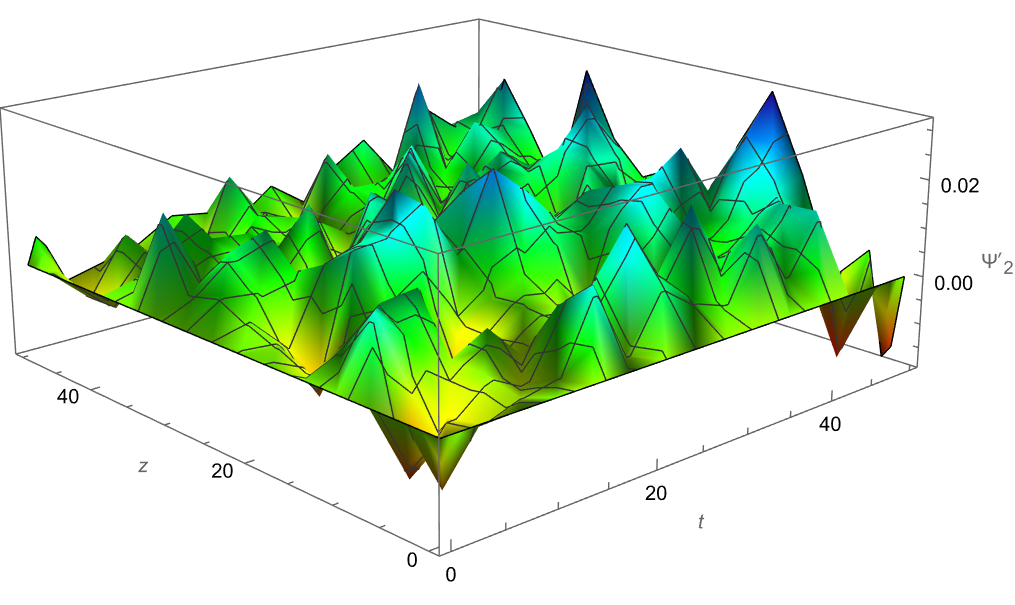}
      \caption{$\Psi'_2(t,z)$}
      \label{fig:psi2allGW}
    \end{subfigure}
    \hfill
    \caption{Figures showing $\Psi'_1(t,z)$ and $\Psi'_2(t,z)$ in the presence of a gravitational wave and condensate to second order in the self-interactions. We choose $g_{ym}=0.5$, $c_1=c_2=1$, $A_+ = 0.0001$, $\omega_g=50$,  $\Psi'_1(0,z)=\Psi'_1(t,0) = \Psi'_2(0,z)=\Psi'_2(t,0) = 0 $ and $\partial_z \Psi'_1(t,0) = \partial_z \Psi'_2(t,0) = 0 $.}
    \label{fig:psiallgw}
\end{figure}

Similarly, in the case of GW, the Eqs. (\ref{eq:psiGW}) \& (\ref{eq:psiGW1}) changed to 
\begin{eqnarray}{\label{eq:psiGWall}}
     \Box \Psi'_1 - 2 g_{ym} U \partial_z \Psi'_2-g_{ym}^2 ({\Psi'_1}^3+\Psi'_1 {\Psi'_2}^2) & = & A_+( g_{ym}^2 U^3 \cos(\omega_g(t-z))+ \partial_0 U A_+ \omega_g \sin(\omega_g(t-z)), \\ 
     \label{eq:psiGWall1}
     \Box \Psi'_2 + 2 g_{ym} U \partial_z \Psi'_1-g_{ym}^2({\Psi'_2}^3+{\Psi'_1}^2 \Psi'_2) & = & -A_+ g_{ym} \omega_g U^2 \sin(\omega_g(t-z)).  
\end{eqnarray}
The numerical solutions for the above equations are shown in Fig. (\ref{fig:psiallgw}). It is very hard to find the analytical solutions for these equations as opposed to the case in the previous section.

We have found that the transverse modes behave differently to all orders in the interactions, with GW. We also observe that a GW interacting with the condensate naturally induces the transverse modes. Thus we are ready to answer the energetics question, can GW induce a decay of the YM condensate? In the next section, we try a different ansatz and obtain a non-perturbative analysis of the `particles' and then analyze the energy of the system. 

\subsection{Summary:} 
As is evident from the above, the GW changes the nature of the fluctuations over a YM condensate introduced in \cite{Prokhorov}. In particular, the transverse modes can be induced, even if the boundary conditions are set to zero for the transverse perturbations. The in-homogeneous terms in the Eqs. (\ref{eq:psiGWall} \& \ref{eq:psiGWall1}), are non-zero at $t=0$, and or $z=0$. This gives rise to non-zero solutions of $\Psi_1',\Psi_2'$, even if $\Psi'_{1(2)}(0)=0$. This illustrates that a GW can induce these `plasmon' modes from the condensate. This should then cause the condensate to lose energy into the `particles' as in \cite{Prokhorov}. The longitudinal modes of the analysis of \cite{Prokhorov} do not see the GW. However, we try to analyse this split into transverse longitudinal vector modes in a more transparent way in the following section to analyze the energetics.

\section{Vector Decomposition of YM Gauge Field}{\label{sec:3}}
In this section, we discuss a different decomposition of the YM vector than what is used in \cite{Prokhorov}. This is motivated by \cite{Prokhorov}, but we have a vector fluctuation of the condensate. This discussion avoids the use of the triads and provides an independent analysis of the YM condensate, though very much in spirit to \cite{Prokhorov}. The triads are useful to introduce gauge-invariant fluctuations, but not to identify the longitudinal and transverse vector modes of the YM gauge field.
We use the following decomposition
\begin{equation}
    A_i^a(t,\mathbf{x})=  U(t) \delta^a_i +n_i \Phi^a (t,\mathbf{x}) + \epsilon_{ijk} n_j s_k ^{\sigma} \chi^a_{\sigma}(t,\mathbf{x}).
    \label{eqn:vectdecomp}
\end{equation}

Here, we have assumed, using the vector decomposition that it has longitudinal modes $\Phi^a$ for a wave propagating in the direction $\mathbf{n}$. These are three, in number, in principle, but we set $\mathbf{n}\cdot \mathbold{\Phi}=0$, which makes the number of independent components as 2.  The $\chi_{\sigma}^a$ represents the transverse mode to $\mathbf{n}$ and these are six in number. (There are two transverse directions to $\mathbf{n}$ and that is obtained using two arbitrary vectors $\mathbf{s}_{\sigma}$, such that $\mathbf{n}\times \mathbf{s}^{\sigma}$ identifies a vector perpendicular to $\mathbf{n}$.)
If we see the above decomposition then we find that the total modes would be: $U(t) \rightarrow 1, \Phi^a \rightarrow 3, \chi_{\sigma}^a \rightarrow 6$ which is 10.
Note the $U(t)$ cannot be classified as a longitudinal or transverse component. It is the trace of the Gauge field, $A_i^i=3 U(t)$.  This trace is implemented using the soldering triads $e^i_a=\delta^i_a$ as it contracts internal indices and the space indices of the gauge field. 
To identify the $\Phi^a$ and $\chi^a_{\sigma}$ as independent of U(t) or the internal trace components, the following conditions are required:
\begin{align}
    \mathbf{n} \cdot \mathbold{\Phi} = & 0\\
    \mathbf{n} \cdot \mathbold{\chi}_{\sigma} = & 0.
\end{align}
As such these are modes of the same gauge field, and therefore independence from each other makes sense at the level of the kinetic term, which are de-coupled for these modes.
The gauge condition is still Hamilton's gauge, with the $A_0^a=0$. As the Lagrangian is independent of any space derivative for $\Phi^a$ (which can be attributed to the above condition), we assume it is only dependent on $t$. We also impose another restriction, in the internal directions:

\begin{equation}
    (\mathbf{n}\times \mathbf{s}^\sigma) \cdot \mathbold{\chi}_\sigma=0
\end{equation}
which is to remove the trace contribution from the $\chi^a_{\sigma}$ modes. This reduced another 1 degree of freedom for the $\chi^a_\sigma$, which now has 3 remaining degrees.
The total degrees of freedom of the gauge field is therefore 1+2+3=6. {\em Thus requiring an isotropic `independent'  condensate reduces the number of degrees of freedom of the YM vector from 9 to 6.} Now, of the above conditions, the first two are required to decouple the $U(t)$ kinetic terms from the plasmon modes. This is an important observation, as an isotropic and homogeneous condensate cannot be naturally obtained as dynamically decoupled from the gauge fields unless some additional conditions are imposed. This in particular the same as the \cite{Prokhorov} assumption of the gluon Heisenberg state.

Next, considering the YM wave propagates in the z-direction i.e. $\mathbf{n}=(0,0,1)$; $\mathbf{s}^1=(1,0,0)$, $\mathbf{s}^2=(0,1,0)$ and taking $\chi_{1}^2= \chi_{2}^1=0$, (this is for a specific solution choice) we derived the equations of motion for $U(t)$, $\Phi^{1} = \Phi_1, \Phi^2 = \Phi_2$, $\chi^1_1=\chi_1$ and $\chi^2_2=\chi_2$. The details regarding the derivation of equations of motion are given in Appendix \ref{AppA}.

\begin{align}{\label{eq:unogw}}
  \partial_t^2 U - \frac{2g_{ym}}{3} U  \partial_z (\chi_1 + \chi_2) + g_{ym}^2\Big[ 2 U^3 + \frac{1}{3} U(\Phi_1^2+ \Phi_2^2)  +  \frac{1}{3} U(\chi_1+\chi_2)^2 - \frac{1}{3} (\chi_1-\chi_2) \Phi_1\Phi_2\Big] &=0,\\
  {\label{eq:phi1}}
  \partial_t^2 \Phi_1 + g_{ym}^2 \Big[U^2 \Phi_1 - U (\chi_1-\chi_2) \Phi_2 +  \Phi_1 \chi_2^2\Big] &=0,\\
  {\label{eq:phi2}}
  \partial_t^2 \Phi_2 + g_{ym}^2 \Big[ U^2 \Phi_2 -  U (\chi_1-\chi_2) \Phi_1 + \Phi_2 \chi_1^2\Big] &=0,\\
  {\label{eq:chi1}}
  \Box\chi_1 + g_{ym}^2 \Big[ U\Phi_1 \Phi_2 -  U^2 (\chi_1+\chi_2) -  \chi_1\chi_2^2 -  \Phi_2^2 \chi_1\Big] &=0,\\
  {\label{eq:chi2}}
  \Box\chi_2 - g_{ym}^2 \Big[ U\Phi_1 \Phi_2 +  U^2 (\chi_1+\chi_2) +  \chi^2_1\chi_2 +  \Phi_1^2 \chi_2\Big] &=0.
\end{align}
The above is derived from Yang-Mills equations in vacuum for the vector modes identified in 
Eq. (\ref{eqn:vectdecomp}). With the use of the constraints discussed above, the overall degrees of freedom are $1 (U(t)) + 2 (\Phi^a) + 3 (\chi^a_{\sigma})=6$, which is the usual count for a massless $SU(2)$ gauge field.
We have also set some terms to zero which are of the form $\mathbf{n} \times \mathbold{\nabla} \chi^a_{\sigma}=0$. This follows from the fact that the modes are propagating in the direction specified by $\mathbf{n}$.

Note that the Gauss constraint is automatically satisfied when the above equations are solved for the $U(t), \Phi_1,\Phi_2,\chi_1,\chi_2$. As there are no derivatives in space for $\Phi_1,\Phi_2$, we can take them to be dependent on time only without losing any dynamical information. All the space-dependent dynamics therefore are in the transverse modes, and we study that in the section with the GW. 
To facilitate a non-perturbative analysis of energy exchange we assume in the first approximation, that all the modes are independent of space, and are functions of time only. This allows us to describe the exchange of energy between the condensate and the longitudinal and transverse modes to all orders in a clean way.  

\begin{figure}[ht]
    \centering
    \begin{subfigure}{0.4\textwidth}
      \includegraphics[width=6cm,height=4.5cm]{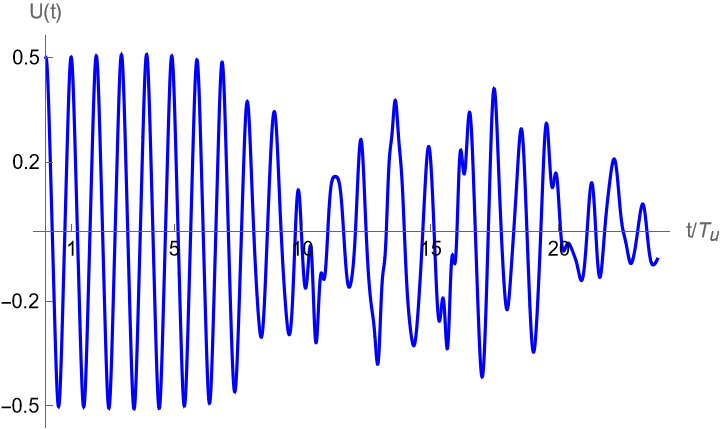}
      \caption{$U(t)$}
      \label{fig:U}
    \end{subfigure}
    \begin{subfigure}{0.4\textwidth}
      \includegraphics[width=6cm,height=4.5cm]{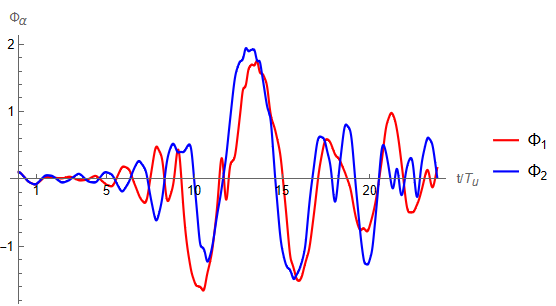}
      \caption{$\Phi_1(t)\ \&\ \Phi_2(t)$}
      \label{fig:phi12}
    \end{subfigure}
    \begin{subfigure}{0.6\textwidth}
    \centering
    \includegraphics[width=6.5cm, height=5cm]{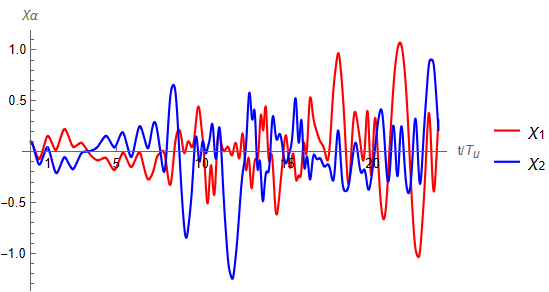}
    \caption{$\chi_1(t)$ \& $\chi_2(t)$}
    \label{fig:chi12}
    \end{subfigure}
    \caption{Figures showing $U$, $\Phi_1$, $\Phi_2$, $\chi_1$ and $\chi_2$ as a function of time. We choose $g_{ym}=0.5$, $U(0)=0.5$, $\Phi_1(0)=0.1$, $\Phi_2(0)=0.1$, $\chi_1(0)=0.1$, $\chi_2(0)=0.1$, $\dot{U}(0)=0$, $\dot{\Phi}_1(0)=0$, $\dot{\Phi}_2(0)=0$, $\dot{\chi}_1(0)=0$ and $\dot{\chi}_2(0)=0$, where $\dot{} \equiv d/dt$.}
    \label{fig:vecsol}
\end{figure}

We solved the above-coupled equations numerically. The time axis of the graphs was given in terms of the period of unperturbed condensate solution ($c_1 {\rm sn}(g_{ym}c_1(t+c_2),-1)$) which is $T_u=(4 K(-1))/(g_{ym} c_1)$, where $K(-1)$ is the complete elliptic function. To keep all the information of the Jacobi function in $c_1$, we choose the initial conditions such that $U(0)=c_1$ and $\dot{U}(0)=0$ which gives the $c_2 = K(-1)/(g_{ym} c_1) $. Then, the time period of unperturbed condensate solution is $T_u=(4 K(-1))/(g_{ym} U(0))$. We find that the amplitude of condensate starts decreasing only after a certain time, and then instead of a continuous decay, it increases again. This is shown in Fig. (\ref{fig:U}). In the quantum scenario, this late perturbative decay of condensate is related to the vacuum expectation value of the condensate \cite{delaydecay}. The time at which the decay of condensate starts is defined as delay decay time ($T_0$). Also, one can see from Eqs. (\ref{eq:phi1}) \& (\ref{eq:phi2}), the $\Phi_1$ and $\Phi_2$ have almost the same functional behavior with a difference in quadratic dependence on $\chi_2^2$ and $\chi_1^2$, respectively. From Eqs. (\ref{eq:chi1}) \& (\ref{eq:chi2}), the transverse modes differential equations differ from one another with signs which results in the almost complementary behavior of these functions as shown in Fig. (\ref{fig:chi12}). From Fig. (\ref{fig:vecsol}), we see that the longitudinal modes reach a very high magnitude as compared to the transverse modes for the same amount of time. This is again due to the quadratic nature of the potential terms in the Eqs. (\ref{eq:phi1} \& \ref{eq:phi2}).


\begin{figure}[htbp]
    \centering
    \includegraphics[scale=0.8]{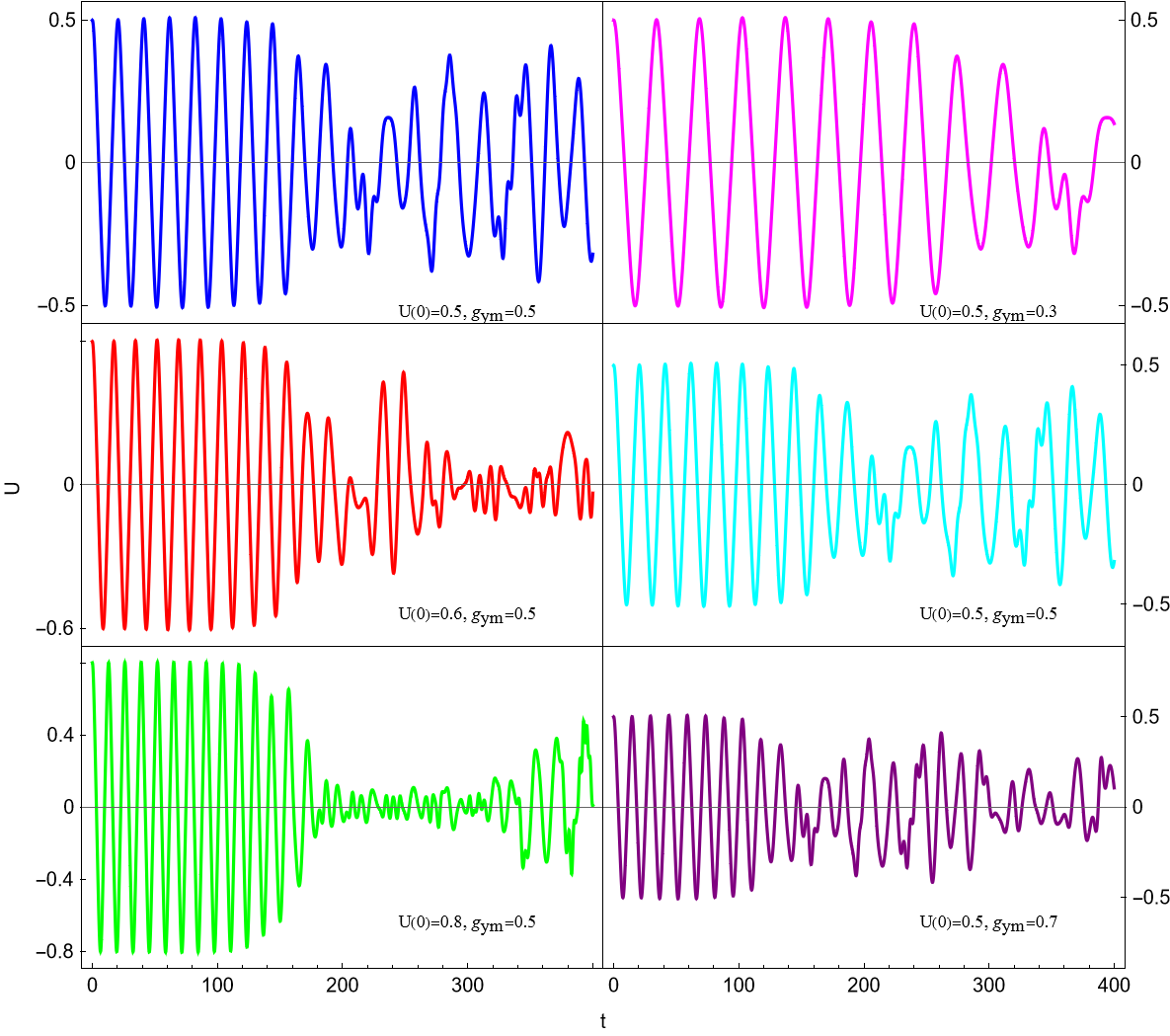}
    \caption{$U$ for different initial conditions and different coupling constants. We choose $\Phi_1(0)=0.1$, $\Phi_2(0)=0.1$,$\chi_1(0)=0.1$, $\chi_2(0)=0.1$, $\dot{\Phi}_1(0)=0$, $\dot{\Phi}_2(0)=0$,  $\dot{U}(0)=0$, $\dot{\chi}_1(0)=0$ and $\dot{\chi}_2(0)=0$.}
    \label{fig:Udifficg}
\end{figure}


If we change the initial value of condensate or coupling constant, we find that the point ($T_0$) where the condensate amplitude starts decaying also changes. This is shown in Fig. (\ref{fig:Udifficg}). If we keep increasing the initial value of condensate while keeping the coupling constant the same, the delay decay time decreases. If we keep the initial value of condensate to be constant while increasing the coupling constant, then also the delay decay time decreases. Since the Eqs. (\ref{eq:chi1}) \& (\ref{eq:chi2}) are coupled, one can see that either one of the transverse modes will be generated even if the other mode is zero. Also, these Eqs. (\ref{eq:chi1}) \& (\ref{eq:chi2}) contain a term ($g_{ym}^2 U \Phi_1 \Phi_2$) independent of transverse modes, this means we can generate the transverse modes even if they were zero initially as long as condensate and longitudinal modes are nonzero.


\textbf{Energy Analysis}

For the energy analysis, it is useful to construct the Hamiltonian density ($\mathcal{H}$) of the SU(2) YM system as a sum of contributions from condensate ($\mathcal{H}_u$), particles ($\mathcal{H}_p$) and the interaction term ($\mathcal{H}_{int}$) as follows

\begin{align} {\label{eq:Htotal}}
    \mathcal{H} &= \mathcal{H}_u +\mathcal{H}_p +\mathcal{H}_{int},\; \mathrm{where}\\
    \mathcal{H}_u &= \frac{3}{2} (\p_t U \p_t U +g_{ym}^2 U^4),\\
    \mathcal{H}_p &= \frac{1}{2} \left( (\p_t \Phi_1)^2 +(\p_t \Phi_2)^2 + (\p_t \chi_1)^2 + (\p_t \chi_2)^2 +g_{ym}^2 \chi_1^2 \chi_2^2 \right),\\
    \mathcal{H}_{int} &= \frac{1}{2} g_{ym}^2 \left[U^2 \left( \Phi_1^2 + \Phi_2^2 +\chi_1^2 +\chi_2^2 \right) + \Phi_1^2 \chi_2^2 + \Phi_2^2 \chi_1^2 +2 U^2 \chi_1 \chi_2 -2 U \Phi_1 \Phi_2 (\chi_1 - \chi_2 ) \right].
\end{align}

\begin{figure}[htbp]
    \centering
    \includegraphics[scale=0.43]{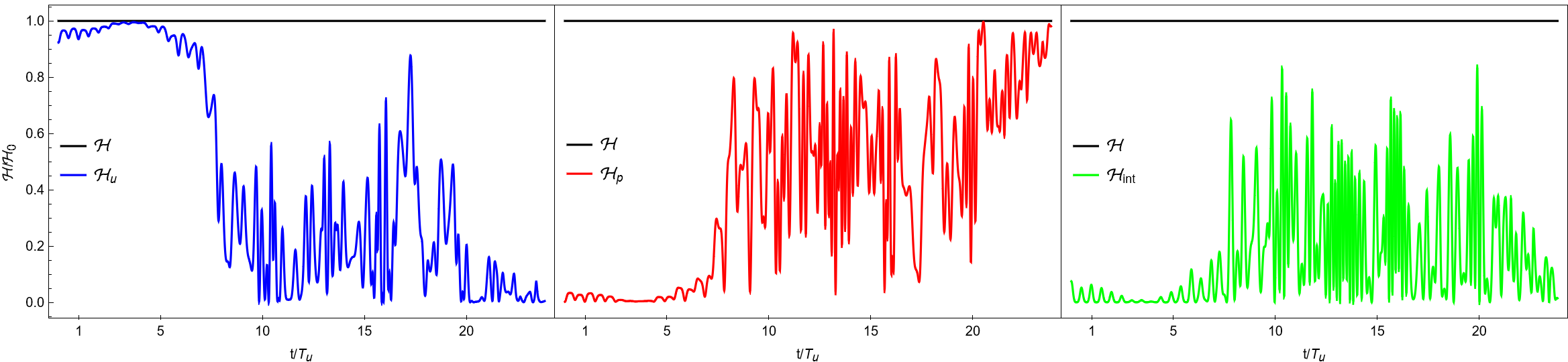}
    \caption{Figure showing the contributions of the condensate $\mathcal{H}_u(t)$, YM wave modes $\mathcal{H}_p(t)$ and interaction terms $\mathcal{H}_{int}(t)$ to the total energy $\mathcal{H}(t)$. We use the same initial conditions as in Fig. (\ref{fig:vecsol}).} 
    \label{fig:Hupint}
\end{figure}

We see that the energy of the condensate ($ \mathcal{H}_u$) is exchanged with the other modes ($\mathcal{H}_p$), but it is oscillatory. After $T_0$ time, we see that there is an energy swap effect between condensate and modes as predicted in \cite{Prokhorov}. This is shown in Fig.(\ref{fig:Hupint}). 

\subsection{In the presence of Gravitational Waves}
Next, we study the above system in the presence of gravitational waves. Consider a gravitational wave propagating in the $z$-direction and assuming only one polarization, we can write the GW as $h_+(t,z)=A_+ \cos(\omega_g (t-z))$. Then, the equations of motion are as below:

\begin{align}
  \partial_t^2 U - \frac{2g_{ym}}{3} U (\partial_z (\chi_1 + \chi_2) +h_+ \partial_z (\chi_1 - \chi_2)) + g_{ym}^2\Big[ 2 U^3 + \frac{1}{3} U(\Phi_1^2+ \Phi_2^2)  +  \frac{1}{3} U(\chi_1+\chi_2)^2 \nonumber \\  - \frac{1}{3} (\chi_1-\chi_2) \Phi_1\Phi_2\Big] -\frac{g_{ym}^2 h_+}{3} \Big[U \left( \chi_2^2-\chi_1^2 + \Phi_2^2-\Phi_1^2\right) + \Phi_1 \Phi_2 (\chi_1+\chi_2)\Big]=0 ,\\
  \label{eq:phi12GW}
  \partial_t^2 \Phi_1 + g_{ym}^2 \Big[U^2 \Phi_1 - U (\chi_1-\chi_2) \Phi_2 +  \Phi_1 \chi_2^2\Big]-g_{ym}^2 h_+\left[\chi_2^2 \Phi_1 - U^2 \Phi_1+ \Phi_2 (\chi_1+\chi_2) U\right]=0, \\
  \label{eq:phi23GW}
   \partial_t^2 \Phi_2 + g_{ym}^2 \Big[ U^2 \Phi_2 -  U (\chi_1-\chi_2) \Phi_1 + \Phi_2 \chi_1^2 \Big]-g_{ym}^2 h_+\left[-\chi_1^2 \Phi_2 + U^2 \Phi_2+ \Phi_1 (\chi_1+\chi_2) U\right]=0, \\
   \label{eq:chi1GW}
  \Box\chi_1 +h_+ \Box \chi_1 - \p_t h_+ \p_t \chi_1 +\p_z h_+ \p_z \chi_1 - g_{ym} U^2 \p_z h_+ + g_{ym}^2 \Big[ U\Phi_1 \Phi_2 -  U^2 (\chi_1+\chi_2) -  \chi_1\chi_2^2 \nonumber\\ +  \Phi_2^2 \chi_1\Big] + g_{ym}^2h_+ \left[-\chi_1 \Phi_2^2 - U^2 \chi_1 + \Phi_1 \Phi_2 U\right] =0,\\
  \label{eq:chi2GW}
  \Box\chi_2 -h_+ \Box \chi_2 + \p_t h_+ \p_t \chi_2 -\p_z h_+ \p_z \chi_2 +g_{ym} U^2 \p_z h_+ - g_{ym}^2 \Big[ U\Phi_1 \Phi_2 +  U^2 (\chi_1+\chi_2) +  \chi^2_1\chi_2 \nonumber\\ +  \Phi_1^2 \chi_2\Big] + g_{ym}^2h_+ \left[\chi_2 \Phi_1^2 + U^2 \chi_2 + \Phi_1 \Phi_2 U\right]=0.
\end{align}

\begin{figure}[htbp]
    \centering
    \begin{subfigure}{0.4\textwidth}
      \includegraphics[width=6cm,height=4.5cm]{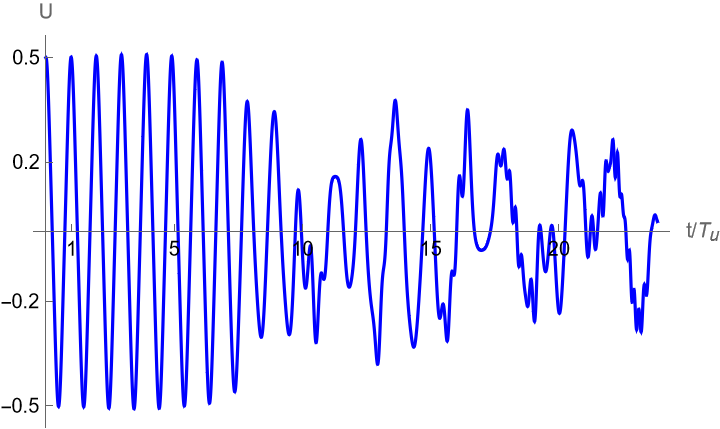}
      \caption{$U(t)$}
      \label{fig:Ugw}
    \end{subfigure}
    \begin{subfigure}{0.4\textwidth}
      \includegraphics[width=6cm,height=4.5cm]{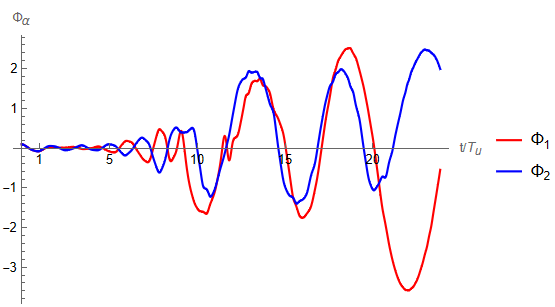}
      \caption{$\Phi_1(t)\ \&\ \Phi_2(t)$}
      \label{fig:phi12gw}
    \end{subfigure}
    \begin{subfigure}{0.6\textwidth}
    \centering
    \includegraphics[width=6.5cm, height=5cm]{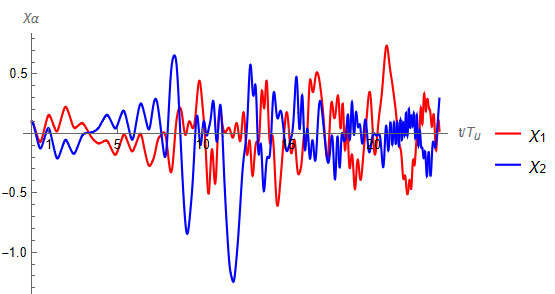}
    \caption{$\chi_1(t)$ \& $\chi_2(t)$}
    \label{fig:chi12gw}
    \end{subfigure}
    \caption{Figures showing $U$, $\Phi_1$, $\Phi_2$, $\chi_1$, and $\chi_2$ as a function of time in the presence of a gravitational wave. We choose $g_{ym}=0.5$, $A_+=0.01$, $\omega_g=100$, $U(0)=0.5$, $\Phi_1(0)=0.1$, $\Phi_2(0)=0.1$, $\chi_1(0)=0.1$, $\chi_2(0)=0.1$, $\dot{U}(0)=0$, $\dot{\Phi}_1(0)=0$, $\dot{\Phi}_2(0)=0$, $\dot{\chi}_1(0)=0$ and $\dot{\chi}_2(0)=0$ .}
    \label{fig:vecsolgw}
\end{figure}

\begin{figure}
    \centering
    \begin{subfigure}{0.45\textwidth}
    \includegraphics[width=7.5cm,height=5cm]{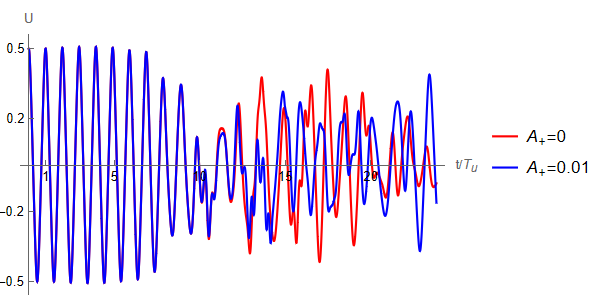}
    \caption{$U(t)$ without and with GW case}
    \end{subfigure}
    \hfill
    \begin{subfigure}{0.45\textwidth}
      \includegraphics[width=7.5cm,height=5cm]{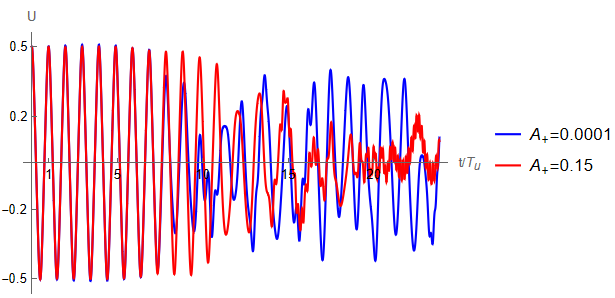}
      \caption{$U(t)$ with different GW amplitudes}
    \end{subfigure}
    \caption{Figure showing $U(t)$ in different scenarios. We choose the other initial conditions as same as in Fig. (\ref{fig:vecsolgw}).}
    \label{fig:ugwcomp}
\end{figure}

To begin with, we set $z=0$ and the equations are solved numerically considering the functions depend only on time, and plots are shown in Fig. (\ref{fig:vecsolgw}). One can see from Fig. (\ref{fig:ugwcomp}) that the solutions with GW approach the solutions without GW if the amplitude of GW ($A_+$) is much smaller than that of condensate and modes. We also find that if we take the GW amplitude ($A_+$) to be as high as that of condensate, the condensate solution tends to become stable, even if the other modes are generated. From Eqs. (\ref{eq:chi1GW}) \& (\ref{eq:chi2GW}), there is a term ($g_{ym} U^2 \partial_z h_+$) which does not depend on any wave modes. This shows that we can generate transverse modes even if they were zero initially. This makes sense if we think that the GW initiates the decay of the condensate.

Similarly, we construct the Hamiltonian density ($\mathcal{H}_{GW}$) in the presence of GWs as follows
\begin{align}
    \mathcal{H}_{GW} ={} & \mathcal{H}_{uGW} +\mathcal{H}_{pGW} +\mathcal{H}_{intGW},\; \mathrm{where}\\
    \mathcal{H}_{uGW} = {}& \frac{3}{2} (\p_t U \p_t U +g_{ym}^2 U^4),\\
    \mathcal{H}_{pGW} = {}& \frac{1}{2} \left( (\p_t \Phi_1)^2 +(\p_t \Phi_2)^2 + (\p_t \chi_1)^2 + (\p_t \chi_2)^2 +g_{ym}^2 \chi_1^2 \chi_2^2 \right), \\
    \mathcal{H}_{intGW} = {}& \frac{1}{2} g_{ym}^2 \left[ U^2 \left( \Phi_1^2 + \Phi_2^2 +\chi_1^2 +\chi_2^2 \right) + h_+ \left( \Phi_1^2 - \Phi_2^2 +\chi_1^2 -\chi_2^2 \right) \right. \nonumber \\
     & {} \left.+ (1-h_+) \Phi_1^2 \chi_2^2 + (1+h_+) \Phi_2^2 \chi_1^2 + 2 U^2 \chi_1 \chi_2 \right. \nonumber \\ 
     & {} \left. -2 U \Phi_1 \Phi_2 ((1+h_+)\chi_1 - (1-h_+) \chi_2 ) \right] 
     +\frac{1}{2} h_+ \left( (\p_t \chi_1)^2 - (\p_t \chi_2)^2  \right).\label{eqn:intener}
\end{align}

\begin{figure}[htbp]
    \centering
    \begin{subfigure}{0.4\textwidth}
      \includegraphics[width=6.5cm,height=4cm]{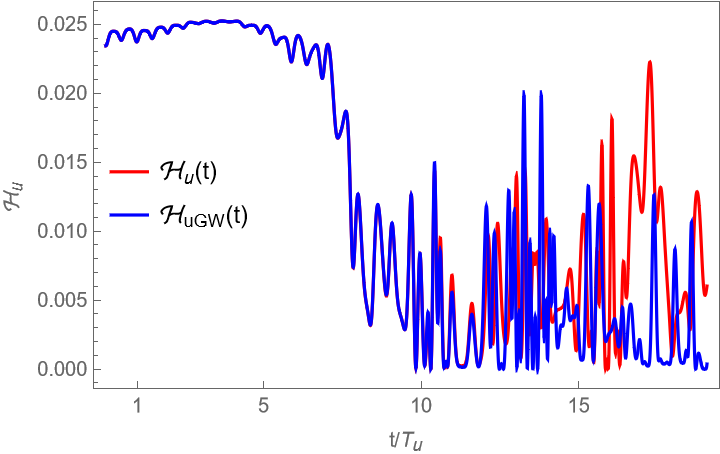}
    \end{subfigure}
    \begin{subfigure}{0.4\textwidth}
      \includegraphics[width=6.5cm,height=4cm]{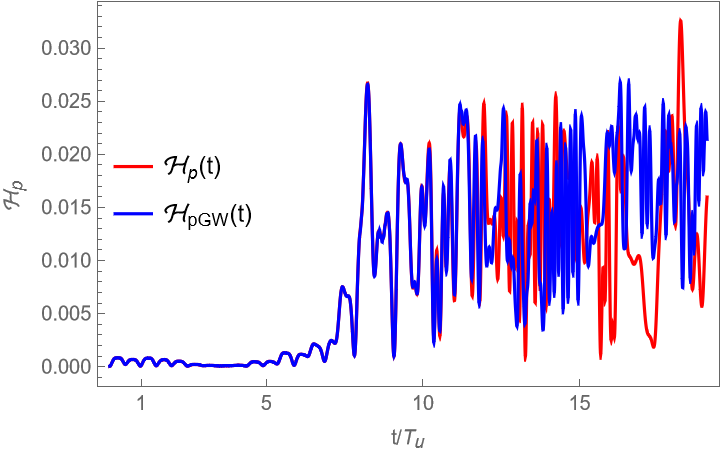}
    \end{subfigure}
    \caption{Figure showing energy density of condensate ($\mathcal{H}_u$) and particles/modes ($\mathcal{H}_p$) with and without GW cases. We choose $A_+ =0.01$, $\omega_g =100$ and same initial conditions as used in Fig. (\ref{fig:vecsolgw}).}
    \label{fig:Eucomp}
\end{figure}

\begin{figure}[htbp]
    \centering
    \begin{subfigure}{0.45\textwidth}
      \includegraphics[width=6cm,height=4cm]{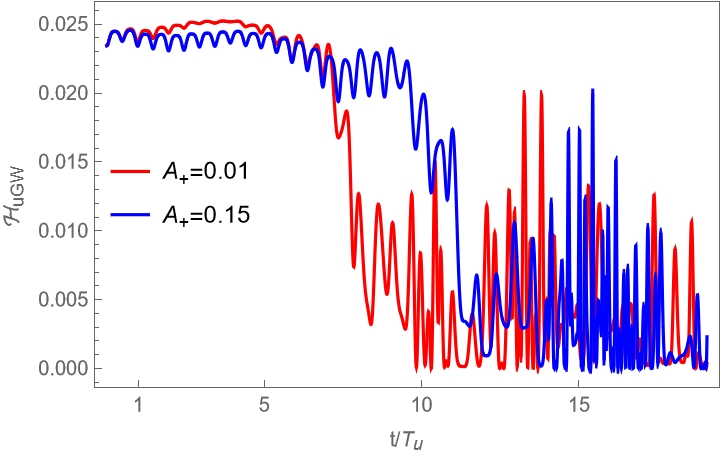}
      \caption{Energy density of condensate ($\mathcal{H}_{uGW}$)}
    \end{subfigure}
    \begin{subfigure}{0.45\textwidth}
      \includegraphics[width=6cm,height=4cm]{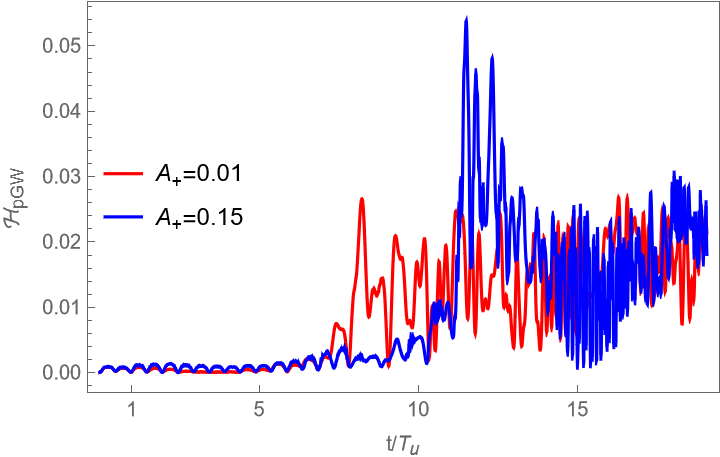}
      \caption{Energy density of the particles $\mathcal{H}_{pGW}$}
    \end{subfigure}
    \caption{Figures showing the impact of high magnitude and high-frequency GWs on condensate decay. We choose the same initial conditions as in Fig. (\ref{fig:vecsolgw}).}
    \label{fig:Eucompgw}
\end{figure}

\begin{figure}[htbp]
    \centering
      \includegraphics[width=6.5cm,height=6cm]{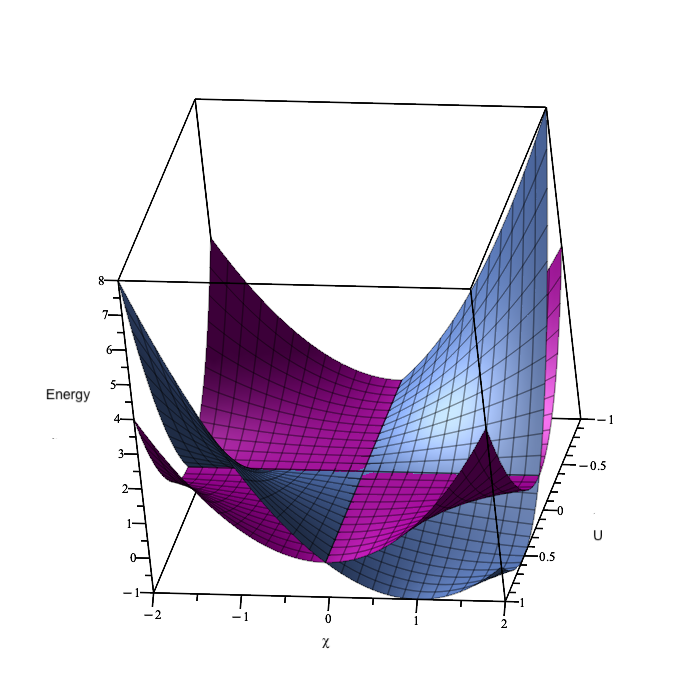}
    \caption{Schematic plot of interaction potential without GW (maroon) and potential with GW (azure), $A_+=0.0001$. In here the potential without GW is $U^2 \chi^2$, where $\chi=\chi_1+\chi_2$, and with the GW it is of the form $U^2 \chi^2 - 2(0.0001)U \chi$. Note we have plotted only the terms that have $U, \chi$ in them and set $\Phi_1=\Phi_2=1$.}
    \label{fig:potential}
\end{figure}

Even though the expressions $\mathcal{H}_{uGW}$ and $\mathcal{H}_{pGW}$ look the same as $\mathcal{H}_{u}$ and $\mathcal{H}_{p}$ in Eqs. (\ref{eq:Htotal}), the values of the expressions are different since the formulas for condensate and modes are different. Thus, we state the two expressions separately. Since the GWs have a very small magnitude, the effect of GWs on the wave modes energy is also very small. We obtained the same energy swap effect as in the previous section (Fig. (\ref{fig:Eucomp})). If we take the GW amplitude to be high or comparable to that of condensate or modes, then GW stabilizes the condensate and the energy swap effect starts at much later times. This is shown in Fig. (\ref{fig:Eucompgw}). This can be attributed to the fact that the interaction term of the condensate with the GW arises due to the other plasmons, whose interaction energy now gets distributed between the condensate and the GW. Therefore the transfer of the condensate energy to the plasmons happens at a delayed pace. Technically there is an interaction of the condensate and the plasmons induced by the $h_+$ which one can see from the sign in the interaction energy reduces the interaction of the condensate and the plasmons, thus affecting the decay time. We have included a schematic plot (\ref{fig:potential}) of how the interaction of potential energy changes due to the gravitational waves. Around the origin or near $\chi_1=\chi_2\approx 0$, the potential is almost parabolic without the GW, but changes in the presence of a GW. Note this is a schematic analysis only of how the energy of interaction of U and the other modes changes with the introduction of $h_+$. The term which couples U to $h_+$ has all the $\Phi_{1,2}$ and $\chi_{1,2}$ modes as in Eq. (\ref{eqn:intener}). For higher values of $\chi$ fields, the potential acquires almost the same shape. Therefore the effect of switching a GW delays the effect.

So far, we studied the system non-perturbatively by suppressing the $z$-dependence of the GW. To study the system with a $z$-dependence of the GW, we solve for the induced perturbations. We first assume there are no longitudinal modes ($\Phi_1 = \Phi_2 =0$) and consider the reduced system ($U, \chi_1, \chi_2$). Note if we assume this without the GW, we find that there is an exchange of energy between condensate and $\chi$ modes, but there is no net energy swap effect. There is an energy swap effect only when the longitudinal modes are non-zero. However, as we analyze, the GW can induce energy exchange without the longitudinal modes. Now consider a perturbation in YM field as $U(t) \rightarrow U(t)+\tilde{U}(t,z) $, $\chi_1(t)\rightarrow \chi_1(t) +\tilde{\chi}_1(t,z) $ and $\chi_2(t)\rightarrow \chi_2(t)+\tilde{\chi}_2(t,z) $. In the linear order in $h_+$ and the perturbations, the equations become 
\begin{align}
    -3\p_t^2 \tilde{U}+2\p_z^2 \tilde{U} + 3 g_{ym} U (\p_z \tilde{\chi}_1+\p_z \tilde{\chi}_2) +g_{ym}^2 \Big[\tilde{U}(-18 U^2-\chi_1^2-\chi_2^2) -2 \tilde{U} \chi_1 \chi_2 -2 U (\chi_1+\chi_2)(\tilde{\chi}_1+\tilde{\chi}_2)\Big]\nonumber \\
    -  g_{ym} U (\chi_1-\chi_2) \p_z h_+ -g^2 h_+ U (\chi_2^2-\chi_1^2)=0,\\
    \Box \tilde{\chi}_1 - 3 g_{ym} U \p_z\tilde{U} - g_{ym}^2 \Big[ U^2(\tilde{\chi}_1+\tilde{\chi}_2) +2U\tilde{U} (\chi_1+\chi_2) + \tilde{\chi}_1 \chi_2^2 + 2\chi_1 \chi_2  \tilde{\chi}_2 \Big]\nonumber \\ - \p_t \chi_1 \p_t h_+ - h_+ \p_t^2 \chi_1  -g_{ym} U^2 \p_z h_+ -g_{ym}^2 h_+ U^2\chi_1 =0, \label{eqn:perturb1}\\
    \Box \tilde{\chi}_2 - 3 g_{ym} U \p_z\tilde{U} - g_{ym}^2 \Big[ U^2(\tilde{\chi}_1+\tilde{\chi}_2) +2U\tilde{U} (\chi_1+\chi_2) +\tilde{\chi}_2 \chi_1^2  + 2\chi_1 \chi_2 \tilde{\chi}_1 \Big]\nonumber \\ 
    +\p_t \chi_2 \p_t h_+ + h_+ \p_t^2 \chi_2 + g_{ym} U^2 \p_z h_+ +g_{ym}^2 h_+ U^2 \chi_2 =0. \label{eqn:perturb2}
\end{align}

\begin{figure}[ht]
    \centering
    \begin{subfigure}{0.4\textwidth}
      \includegraphics[width=6cm,height=4cm]{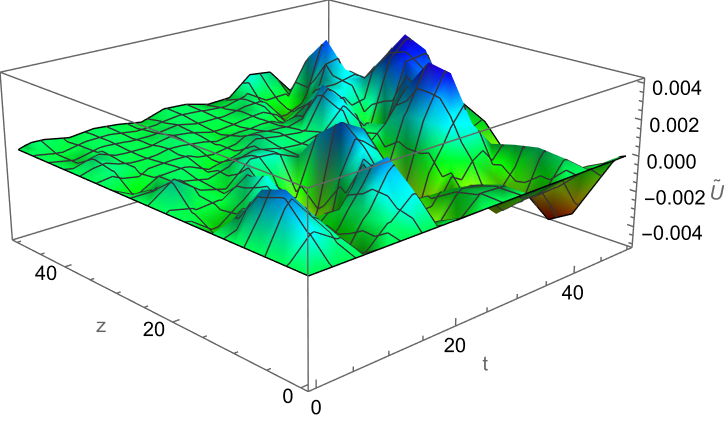}
      \caption{$\tilde{U}(t,z)$}
      \label{fig:Upert}
    \end{subfigure}
    \begin{subfigure}{0.4\textwidth}
      \includegraphics[width=6cm,height=4cm]{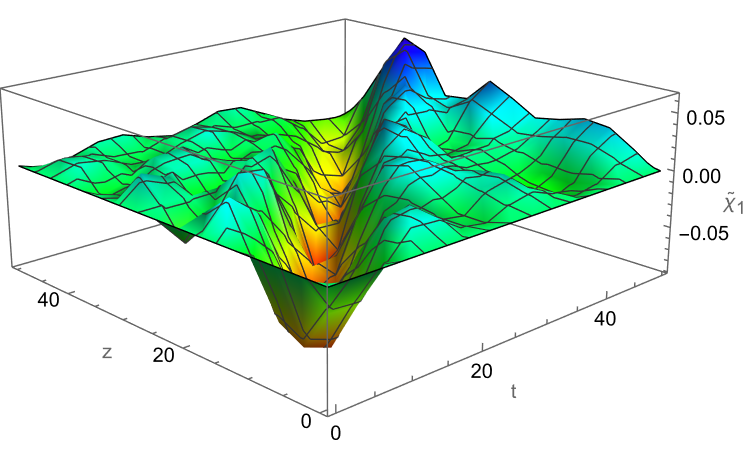}
      \caption{$\tilde{\chi}_1(t,z)$}
      \label{fig:chi1pert}
    \end{subfigure}
    \begin{subfigure}{0.6\textwidth}
    \centering
      \includegraphics[width=6cm,height=4.5cm]{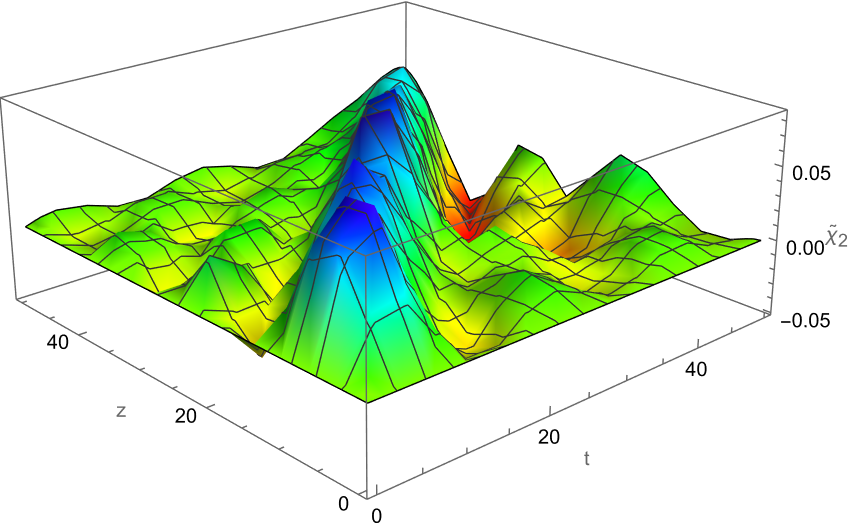}
      \caption{$\tilde{\chi}_2(t,z)$}
      \label{fig:chi2pert}
    \end{subfigure}
    \caption{Figures showing $\tilde{U}(t,z)$, $\tilde{\chi}_1(t,z)$ and $\tilde{\chi}_2(t,z)$ as a function of time and space in the presence of a gravitational wave. We choose $g_{ym}=0.5$, $A_+=0.1$, $\omega_g=100$, $U(0)=1$, $\chi_1(0)=0.1$, $\chi_2(0)=0.2$,$\dot{U}=0$, $\dot{\chi}_1(0)=0$, $\dot{\chi}_2(0)=0$, $\tilde{U}(t,0) = \tilde{U}(0,z) = 0 $, $\tilde{\chi}_1(t,0) = \tilde{\chi}_1(0,z) = 0 $, $\tilde{\chi}_2(t,0) = \tilde{\chi}_2(0,z) = 0 $, $\partial_t \tilde{U}(0,z)= \partial_z \tilde{U}(t,0) = 0 $, $\partial_t \tilde{\chi}_1(0,z)= \partial_z \tilde{\chi}_1(t,0) = 0 $ and $\partial_t \tilde{\chi}_2(0,z)= \partial_z \tilde{\chi}_2(t,0) = 0 $.}
    \label{fig:vecsolgwpert}
\end{figure}

We solved the above equations and the numerical solutions were given in Fig. (\ref{fig:vecsolgwpert}). We can see that the perturbation in $U$ is very small during early times as agreeable with the exact solution (Fig. (\ref{fig:Ugw})). One can see that the magnitude of unperturbed condensate ($\mathcal{O}(1)$) is much greater than the magnitude of perturbed condensate solution ($\mathcal{O}(10^{-3}$)). We also find that the condensate loses its homogeneity and isotropic nature and gets perturbed due to GW.

We also obtained the Hamiltonian density ($\mathcal{H}$) of the reduced system, up to second-order terms, as follows 
\begin{align}
    \mathcal{H} = & \mathcal{H}'+\mathcal{H}''+\mathcal{H}''',\; \mathrm{where}\\
    \mathcal{H}' = & \frac{3}{2} \left[ (\p_t U)^2+g_{ym}^2 U^4 \right]+ \frac{1}{2}\left[ (\p_t \chi_1)^2 +(\p_t \chi_2)^2 + g_{ym}^2 \chi_1^2 \chi_2^2 \right] +\frac12 g_{ym}^2 U^2 (\chi_1+\chi_2)^2,\\
    \mathcal{H}''= & 3 \p_t U \p_t \tilde{U} +\p_t \chi_1 \p_t \tilde{\chi}_1 + \p_t \chi_2 \p_t \tilde{\chi}_2 + g_{ym} \left[ 2U \p_z \tilde{U} (\chi_1+\chi_2)- 2U^2 (\p_z \tilde{\chi}_1 + \p_z \tilde{\chi}_2) \right] \nonumber\\
    & + g_{ym}^2 \left[ 6U^3 \tilde{U} +U^2(\chi_1+\chi_2)(\tilde{\chi}_1+\tilde{\chi}_2) +U \tilde{U}(\chi_1+\chi_2)^2 + \chi_1\chi_2^2 \tilde{\chi}_1 \right]\nonumber\\
    & +\frac12 h_+ ((\p_t \chi_1)^2-(\p_t \chi_2)^2) +\frac12 g_{ym}^2 h_+ U^2 (\chi_1^2-\chi_2^2), \\
     \mathcal{H}'''= & \frac{3}{2} (\p_t \tilde{U})^2 + (\p_z \tilde{U})^2 +\frac{1}{2} \left[ (\p_t \tilde{\chi}_1)^2 +(\p_z \tilde{\chi}_1)^2 +(\p_t \tilde{\chi}_2)^2 + (\p_z \tilde{\chi}_2)^2 \right] + g_{ym} \left[\tilde{U}\p_z \tilde{U} (\chi_1+\chi_2) \right. \nonumber\\
     &  \left. + U \p_z \tilde{U} (\tilde{\chi}_1 + \tilde{\chi}_2 ) - 4 U \tilde{U} (\p_z \tilde{\chi}_1 + \p_z \tilde{\chi}_2 ) \right]  +\frac{1}{2} g_{ym}^2 \left[ 18 U^2 \tilde{U}^2 +\tilde{U}^2 (\chi_1+\chi_2)^2 + 4 U \tilde{U} (\chi_1+\chi_2)  \right. \nonumber \\
     & \left. (\tilde{\chi}_1+\tilde{\chi}_2) +U^2 (\tilde{\chi}_1+\tilde{\chi}_2)^2 + \chi_1^2 \tilde{\chi}_2^2+ \chi_2^2 \tilde{\chi}_1^2 + 4 \chi_1 \chi_2 \tilde{\chi}_1 \tilde{\chi}_2 \right]  + h_+ \left[ \p_t \chi_1 \p_t \tilde{\chi}_1 + \p_t \chi_1 \p_t \tilde{\chi}_2 \right]  \nonumber\\
     &   + g_{ym} h_+ \left( U\p_z \tilde{U} (\chi_1+\chi_2)-2 U^2 (\p_z \tilde{\chi}_1 - \p_z \tilde{\chi}_2 )\right)+\frac12 g_{ym}^2 h_+ \left(2 U \tilde{U} (\chi_1^2-\chi_2^2) + 2 U^2 (\chi_1 \tilde{\chi}_1-\chi_2 \tilde{\chi}_2)\right).
\end{align}

\begin{figure}[htbp]
    \centering
    \begin{subfigure}{0.4\textwidth}
    \centering
      \includegraphics[scale=0.45]{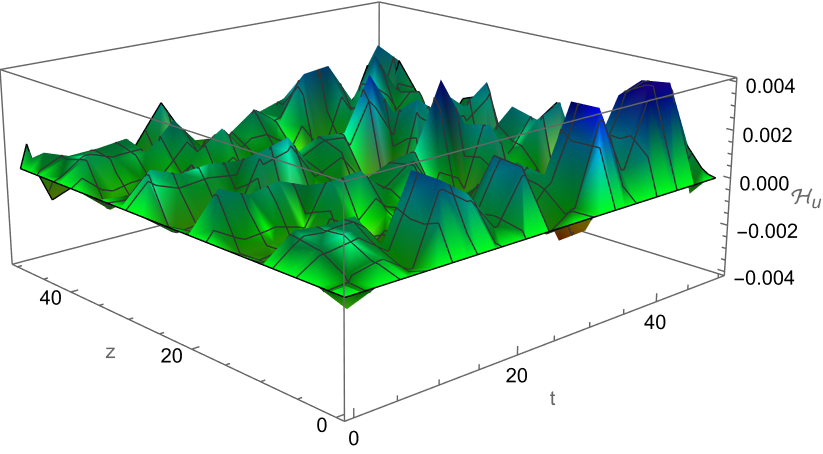}
    \end{subfigure}
    \hfill
    \begin{subfigure}{0.4\textwidth}
     \centering
      \includegraphics[scale=0.45]{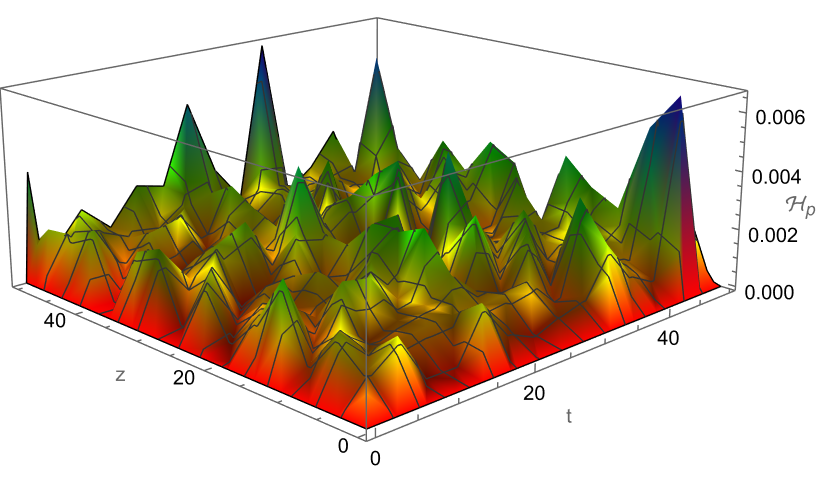}
    \end{subfigure}
    \caption{Figures showing perturbed energy density contribution of condensate (left) and wave modes (right) as a function of time and space in the presence of a gravitational wave. We choose the parameters same as in Fig. (\ref{fig:vecsolgwpert}).}
    \label{fig:Epert}
\end{figure}

\begin{figure}[htbp]
    \begin{subfigure}{0.45\textwidth}
      \includegraphics[scale=0.45]{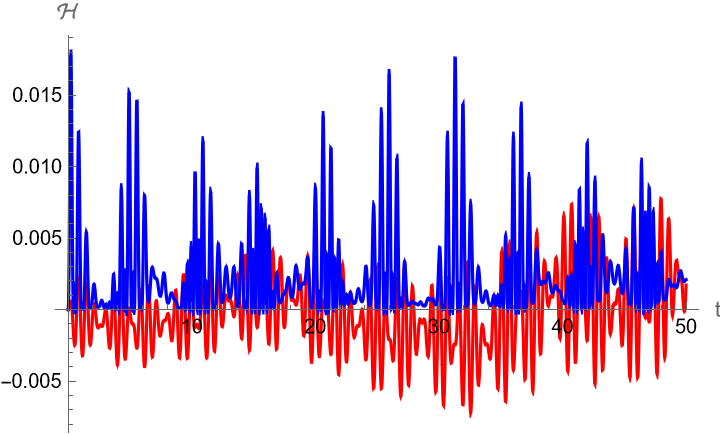}
    \end{subfigure}
    \hfill
    \begin{subfigure}{0.45\textwidth}
      \includegraphics[scale=0.45]{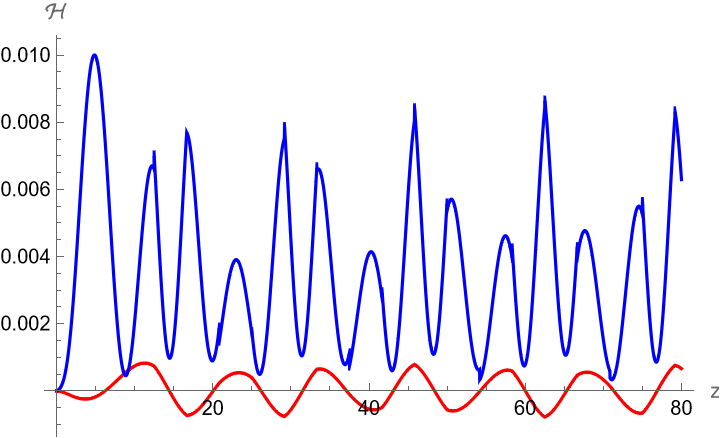}
    \end{subfigure}
    \caption{Figures showing perturbed energy density contribution of condensate (Red plot) and wave modes (Blue plot) as a function of time (at $z=5$) (left) and space (at $t=5$) (right) separately in the presence of a gravitational wave. We choose the parameters same as in Fig. (\ref{fig:vecsolgwpert}).}
    \label{fig:Eperttz}
\end{figure}

The perturbed energy plots are given in Figs. (\ref{fig:Epert}) \& (\ref{fig:Eperttz}) as contributions from condensate and wave modes, respectively. These are the fluctuations over the zero-order energy density contribution of the condensate and modes, so they can be negative as shown in Fig. (\ref{fig:Eperttz}). As one can see the energy contribution due to GW is very small and there is an exchange of energy between condensate and particles but still is oscillatory. If we keep the assumption that the condensate is a homogeneous and isotropic configuration, then the equation for condensate will become a constraint on solutions of $\tilde{\chi}_1$ and $\tilde{\chi}_2$. Thus, we have to perturb the scalar mode with a $z$ dependent function.

\begin{figure}[htbp]
    \begin{subfigure}{0.3\textwidth}
      \includegraphics[scale=0.3]{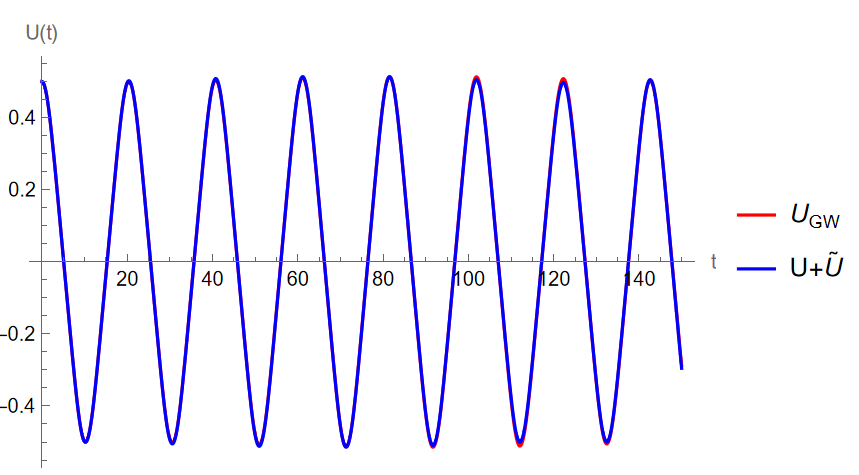}
      \caption{$U(t)$}
      \label{fig:Ucomp}
    \end{subfigure}
    \hfill
    \begin{subfigure}{0.3\textwidth}
      \includegraphics[scale=0.3]{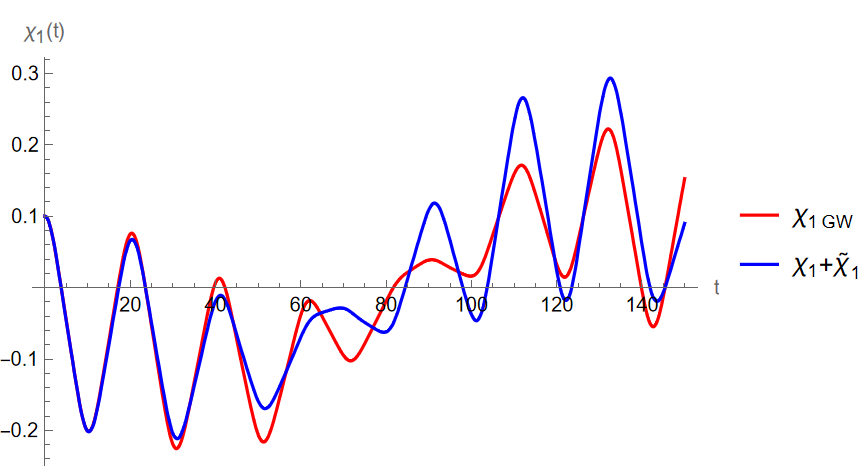}
      \caption{$\chi_1(t)$}
      \label{fig:chi1comp}
    \end{subfigure}
    \hfill
    \begin{subfigure}{0.3\textwidth}
      \includegraphics[scale=0.3]{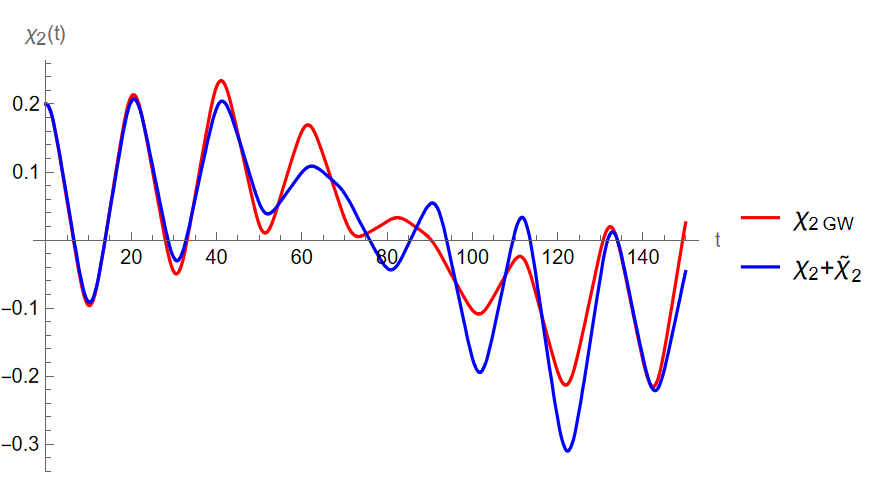}
      \caption{$\chi_2(t)$}
      \label{fig:chi2comp}
    \end{subfigure}
    \caption{Figures showing $U(t)$, $\chi_1(t)$ and $\chi_2(t)$ as a function of time in the presence of a gravitational wave. We choose the parameters same as in Fig.(\ref{fig:vecsolgwpert}). }
    \label{fig:vecsolgwcomp}
\end{figure}

To verify that our perturbation theory works, we compare the solutions obtained Eqs. (\ref{eqn:perturb1} \& \ref{eqn:perturb2}), with the ones obtained non-perturbatively. For comparison with the solutions of Eqs. (\ref{eq:chi1GW}, \ref{eq:chi2GW}) , we consider a reduced system ($U,\chi_1, \chi_2$) by taking only as functions of time (i.e. by solving the system at a particular $z$ value ($z=0$)). Then, we solved the reduced system ($U(t),\chi_1(t), \chi_2(t)$) in two ways: one is finding exact solution ($U_{GW}(t), \chi_{1GW}(t)  \mathrm{and} \chi_{2GW}(t)$), and other is finding the perturbation solution ($\tilde{U}(t),\tilde{\chi}_1(t) \mathrm{and} \tilde{\chi}_2(t) $) by suppressing the $z$-dependence. We plot the solutions in Figs. (\ref{fig:vecsolgwcomp}). As one can see from Fig. (\ref{fig:Ucomp}), the perturbation in $U$ is very tiny and agrees with Fig. (\ref{fig:Upert}). We can restore the longitudinal modes and compute the $z$ dependence of the system with a GW, but the physics of the system has already been obtained here. Restoring the modes should not give any surprises. As one can see from the Eqs. (\ref{eq:phi12GW}) \& (\ref{eq:phi23GW}), there are no z derivatives in $\Phi_1$ and $\Phi_2$. If we perturb the $\Phi_1$ and $\Phi_2$ modes, then the z-dependence on $\Phi_1$ and $\Phi_2$ will be just a Fourier mode dependence with the same frequency as GW.

\subsection{Summary:} In this section, we have taken a vector decomposition of the SU(2) YM vector field. This is motivated by the decomposition of \cite{Prokhorov}, but the `longitudinal' and the `transverse' modes are identified as a set of 2 \& 3 component modes after the restrictions. The Energy exchange of the condensate with the modes is derived numerically, we find that a GW can induce the decay of the condensate into particles, but its effect is to delay the decay even further. Next, we discuss the same model by adding `quarks' or Fermions to the condensate. Note, all our calculations are classical and we have not made a quantum or thermodynamic analysis of the system. However, thermodynamics is based on these fundamental interactions, and thus our study of the interactions can be used for a QGP analysis. Also, our results are valid for SU(2) YM fields but can be extended to SU(3) YM fields.

\section{Dynamics of Fermions in the presence of Yang-Mills Condensate}{\label{sec:4}}
In this section, we have a brief discussion on the dynamics of massless Fermion fields in the presence of YM condensate. The work relating the Fermion production in the presence of $SU(2)$ gauge fields were well studied in \cite{fer2}. We take the $SU(2)$ weak interactions and a lepton doublet for the study to begin with. We take the YM condensate field as obtained in the previous discussion as the Jacobi elliptic function, and try to solve for the Fermion doublet in these backgrounds \cite{pich}. 
The Dirac equation for the Fermions is
\begin{equation}
    i \gamma^{\mu} \partial_{\mu} \psi_{\alpha} + g_{ym} \gamma^{\mu} A_{\mu}^a T^a_{\alpha \beta} \psi_{\beta}=0,
    \label{eq:dirac}
\end{equation}
where $T^a$ are the $SU(2)$ generators, $\gamma^{\mu}$ are the four gamma matrices of the Clifford algebra and $\psi_{\alpha}$ are a $SU(2)$ Fermion doublet with $\alpha=1,2$. Now as the condensate YM field is $A_i^a= U(t) \delta^a_i$, one solves for the Dirac equation in terms of this function. The solutions are analytically obtained in terms of Jacobi elliptic functions. We then study the back reaction of the Fermions on the condensate, in the spirit of \cite{Prokhorov}. Fermions are not classical objects, and it makes sense only to study the bilinears built from them. 
             
Taking the $SU(2)$ generators to be $T^a= \sigma^a/2$, where $\sigma^a $ are the Pauli matrices and labeling the Fermion doublets as ($\psi_1, \psi_2$), the Dirac equation can be written as  
\begin{align}
    i \gamma^{\mu} \partial_{\mu} \psi_1 + \frac{g_{ym}}{2} U(t)\left[\left(\gamma^1-i \gamma^2\right) \psi_2 + \gamma^3 \psi_1\right]=0,\\
    i \gamma^{\mu} \partial_{\mu} \psi_2 + \frac{g_{ym}}{2} U(t)\left[\left(\gamma^1+i \gamma^2\right) \psi_1 - \gamma^3 \psi_2\right]=0.
\end{align}
We further solve these using the chiral representation of the gamma matrices for the Left and Right chirality Fermions as $(\psi_{1L}, \psi_{1R})$ and $(\psi_{2L},\psi_{2R})$, respectively. If the Fermions are quarks and the $SU(2)$ group is the isospin group, then the left quarks are a doublet, whereas the right quarks are singlets. However to begin with we take both the left and right quarks as doublets. Therefore, in principle, these can be a partial $SU(3)$ system.
\begin{align}
    i (\partial_0 - \vec{\sigma}\cdot\vec{\partial})\psi_{1L} -\frac{g_{ym}}{2} U(t)[(\sigma^1-i \sigma^2) \psi_{2L} + \sigma^3 \psi_{1L}]=0,\\
    i (\partial_0 + \vec{\sigma}\cdot\vec{\partial})\psi_{1R} +\frac{g_{ym}}{2} U(t)[(\sigma^1-i \sigma^2) \psi_{2R} + \sigma^3 \psi_{1R}]=0,\\
    i (\partial_0 - \vec{\sigma}\cdot\vec{\partial})\psi_{2L} -\frac{g_{ym}}{2} U(t)[(\sigma^1+i \sigma^2) \psi_{1L} - \sigma^3 \psi_{2L}]=0,\\
    i (\partial_0 + \vec{\sigma}\cdot\vec{\partial})\psi_{2R} +\frac{g_{ym}}{2} U(t)[(\sigma^1+i \sigma^2) \psi_{1R} - \sigma^3 \psi_{2R}]=0.
\end{align}
At the level of the two-component equations for the Weyl Fermions written as ($\psi_{1L1},\psi_{1L2}$) and ($\psi_{2L1},\psi_{2L2}$), we keep only the Left-handed quarks which form a doublet, e.g. ($u_{\rm L}, d_{\rm L} $), as non-zero. In this case, we are taking $SU(2)$ flavour quarks. If the $SU(2)$ quarks were transforming in the subgroup of $SU(3)$, then both the right and left-handed quarks could be doublets.

\begin{figure}[htbp]
    \centering
    \includegraphics[width=0.5\linewidth]{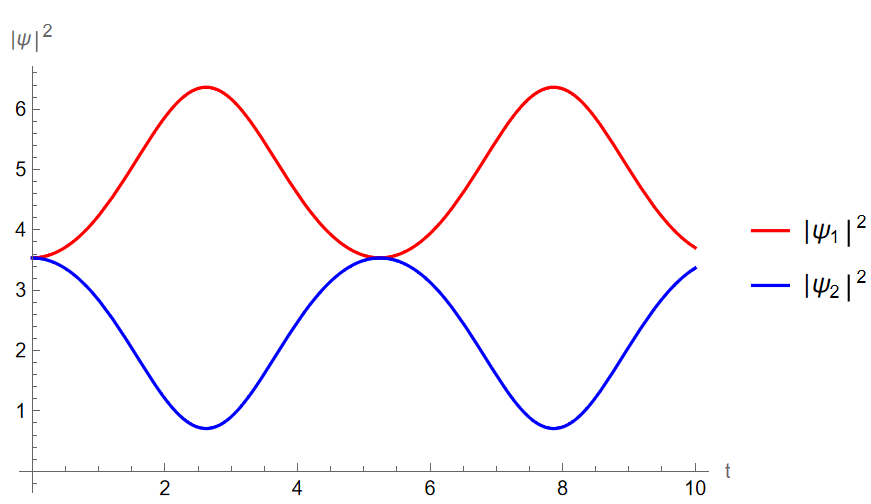}
    \caption{Figure showing the probability densities for $\psi_1$ and $\psi_2$. We use $c_1 = c_3 = c_4 =1$, $A_1= A_2= c_2 =0$ and $g_{ym} = 0.5$.}
    \label{fig:ferprobwithoutGW}
\end{figure}

The equations are solved when the wave functions only depend on time:
\begin{align}
    i \partial_0 \psi_{1L1} - \frac{g_{ym}}{2} U(t) \psi_{1L1}=0, \label{eqn:fermion1}\\
    i \partial_0 \psi_{1L2} - g_{ym} U(t) \psi_{2L1} + \frac{g_{ym}}{2} U(t) \psi_{1L2}=0, \label{eqn:fermion2}\\
    i \partial_0 \psi_{2L1} - g_{ym} U(t) \psi_{1L2} + \frac{g_{ym}}{2} U(t) \psi_{2L1}=0, \label{eqn:fermion3}\\
    i \partial_0 \psi_{2L2} - \frac{g_{ym}}{2} U(t) \psi_{2L2}=0. \label{eqn:fermion4}
\end{align}
As is obvious, the $\psi_{1L1}$ and the $\psi_{2L2}$ are decoupled from each other. If we try to solve for the other wave functions, we get the following solutions:( for a general condensate solution of $ U(t)= c_1 \rm{sn}(g_{ym} c_1(t+c_2), -1)$)
\begin{align}
    \psi_{1L1}(t) &=  A_1 \Lambda^{-1/2},\\
    \psi_{2L2} (t) &= A_2 \Lambda^{-1/2},\\
    \psi_{1L2} (t) &= c_3 \Lambda^{3/2} +c_4 \Lambda^{-1/2},\\
    \psi_{2L1}(t) &= -c_3 \Lambda^{3/2} +c_4 \Lambda^{-1/2},
\end{align}
where $ \Lambda = \left({\rm dn}(g_{ym} c_1(t+c_2), -1)-i \ {\rm cn}(g_{ym} c_1(t+c_2), -1)\right)$, $A_1$, $A_2$, $c_3$ and $c_4$ are integration constants, and $\rm{dn}$ and $\rm{cn}$ are the Jacobi elliptic functions. Out of all constants, only $c_1$ and $c_2$ are real while others are complex constants. One can find the fermion densities for these components ($|\psi_{1L}|^2$ and $|\psi_{2L}|^2$) which tells us about the probability of finding that fermion component in the system. Even though the condensate seems to couple two different flavours ($\psi_{1L2}\ \&\ \psi_{2L1}$) components, the densities generated by these two types of fermions do not mix each other. This indicates that one can always find the system in favour of one flavour only. This is shown in Fig.(\ref{fig:ferprobwithoutGW}). 

Note we eventually discuss a `mixing of fermion flavour' which is obtained by observing the difference $\psi_{1L}^{\dag}\psi_{1L}- \psi^{\dag}_{2L}\psi_{2L}$, which oscillates due to interaction with GW. Here, without the GW, the sign of this quantity remains preserved, if we have $c_3^*c_4$ product as real.\footnote{We thank the referee for pointing to this detail.} This is the case we analyse here. 

If one wants to study the backreaction of fermions on the condensate, one computes the current density using $(j^\mu)^a = \bar{\psi}\gamma^\mu T^a \psi$. For the zero-component of current density ($(j^0)^a$), we find 
\begin{eqnarray}
          (j^0)^1 &= &\frac12(\psi_{1L}^{\dag} \psi_{2L}+ \psi_{2L}^{\dag} \psi_{1L}), \\
          (j^0)^2 &=& -\frac{i}{2} (\psi_{1L}^{\dag} \psi_{2L} - \psi^{\dag}_{2L} \psi_{1L}), \\
          (j^0)^3 &=& \frac12 (|\psi_{1L}|^2 -|\psi_{2L}|^2).
\end{eqnarray}

If one wants to preserve Hamilton's gauge even in the backreaction, we need to set the $\psi_{1L1}=\psi_{2L2}=0$ and $\psi_{1L2}=\psi_{2L1}$ in the ODEs as a boundary condition, then the solution for the first order ODE is
\begin{equation}
         \psi_{1L2}=\psi_{2L1}= A \Lambda^{-1/2},
\end{equation}
where $\Lambda$ is defined before and $|\psi_{1L2}|^2 = |A|^2/\sqrt{2} $. With the choice $\psi_{1L1}=\psi_{2L2}=0$ and $\psi_{1L2}=\psi_{2L1}$, we have Fermion charge density ($(j^0)^a$) to be zero.  One can choose a less restricted condition such as  $\psi_{1L1}=\psi_{2L2}=0$, $\psi_{1L2}\neq 0$, and $\psi_{2L1} \neq 0$. With this choice, we can restrict the gauge field to be in Hamilton's gauge while having a nonzero charge density.


For the space components of the current, we find similar formulas. Putting these in the Yang-Mills equation with a non-zero source term, we can find the back-reaction of the Fermions on the condensate. What is obvious is that the backreaction breaks the isotropy of the system. One cannot have the condensate as $U(t)\delta^a_i$ as the different internal components $a=1,2,3$ will have different sources. Instead of discussing this further, in this article, we will report on this in a separate publication, which will also discuss the quantum interaction \cite{condensate1}.

\subsection{In the presence of a Gravitational wave background}
In this article, we discuss the quarks + condensate system in the background of a gravitational wave \cite{Birrell,collas}. First, we need to define the Dirac equation in curved spacetime as
\begin{equation}{\label{eq:diraceqcurve}}
    i \bar{\gamma}^\mu (\partial_\mu + \Gamma_\mu) \psi_\alpha + g_{ym} \bar{\gamma}^\mu A^a_\mu T^a_{\alpha\beta} \psi_\beta = 0
\end{equation}
where $\bar{\gamma}^\mu$ are the spacetime-dependent gamma matrices and $\Gamma_\mu$ is the spinor affine connection. Considering a +-polarised GW propagating in $z$-direction in the flat background, we solve the Dirac equation. The details regarding the derivation of evaluating the Dirac equation are given in Appendix \ref{App:Dirac}. Since the GW has $z$-dependence, we also get the terms dependent on spatial coordinates. We find that the GW couples to only two of the Fermion wave components. The coupled equations through GW are given below:
\begin{eqnarray} {\label{eq:fermionGW1}}
    i (\partial_0 + \partial_3) \psi_{2L2} + \frac{g_{ym}}{2} h_+ U \psi_{1L1} -\frac{g_{ym}}{2} U \psi_{2L2}&=&0,\\
    \label{eq:fermionGW2}
i (\partial_0 - \partial_3) \psi_{1L1} + \frac{g_{ym}}{2} h_+ U \psi_{2L2} - \frac{g_{ym}}{2}U \psi_{1L1} &=& 0.
\end{eqnarray}

What is obvious is that the two modes in Eqs. (\ref{eqn:fermion1} \& \ref{eqn:fermion4}), are now coupled, due to the GW. As one can see the GW connects the two Fermions from different elements of the $SU(2)$ doublet. We can solve the coupled equations, and the time evolution of the solutions are plotted in the following graphs. It is evident that the Fermion densities fluctuate in time, due to the GW. The flavour transitions induced by a graviton have been studied in \cite{fermionscatter}. The evidence for flavour transitions in a GW background obtained in this paper, are in presence of a condensate. The results will have relevance for QGP and `real' GW interactions with quarks. GWs have been observed, and at these length scales of 200 MeV, which is QCD energy scale, gravity remains classical. Our calculations show that these GW can induce changes in flavour quark fluxes in the presence of a $SU(2)$ condensate. This work might provide interesting insight into the workings of QGP.

\begin{figure}[htbp]
      \centering
      \includegraphics[scale=0.6]{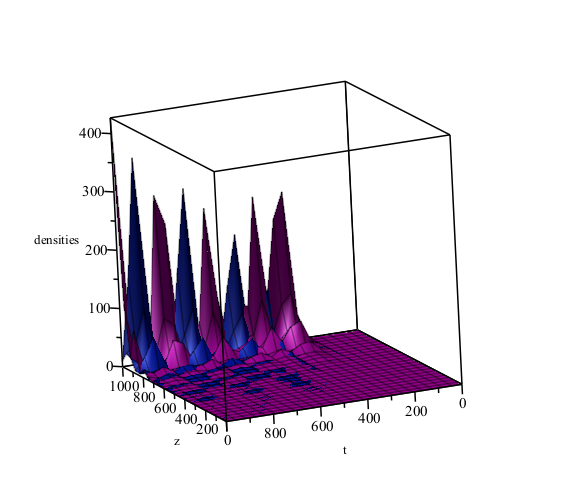}
      \caption{Fermion densities of two flavours with GW at $\omega_g = 10^{-4}, A_+=1$. $|\psi_{1L}|^2$ is plotted in blue color, and the $|\psi_{2L}|^2$ in maroon. In the above, we choose $c_1=1, c_2=1/2, g_{ym}= \sqrt{2}, \psi_{1L1}(t,0) = 10^{-2} U(t)= \psi_{2L2}(t,0), \psi_{1L1}=(0,z)=\psi_{2L2}(0,z)=0$.$\psi_{1L2},\psi_{2L1}$ are set to zero.}
      \label{fig:psi11}
\end{figure}

\begin{figure}[htbp]
     \centering
      \includegraphics[scale=0.4]{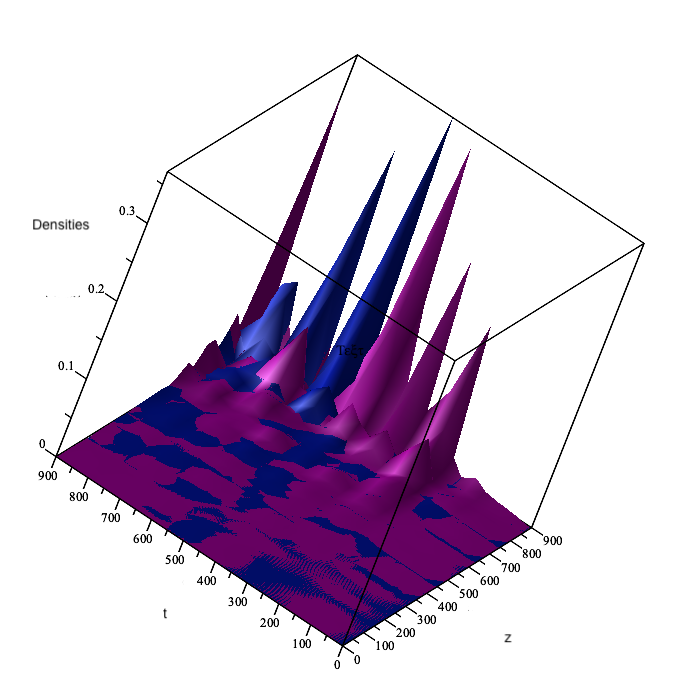}
      \caption{Fermion densities of two different flavours with GW at $\omega_g =10$. $|\psi_{1L}|^2$ is plotted in blue color, and the $|\psi_{2L}|^2$ in maroon. We choose $c_1=1, c_2=1/2, g_{ym}= \sqrt{2}, \psi_{1L1}(t,0) = 10^{-2} U(t)= \psi_{2L2}(t,0), \psi_{1L1}=(0,z)=\psi_{2L2}(0,z)=0$. $\psi_{1L2},\psi_{2L1}$ are set to zero.}
      \label{fig:psi21}
\end{figure}

\begin{figure}[htbp]
    \begin{subfigure}{0.4\textwidth}
      \includegraphics[scale=0.4]{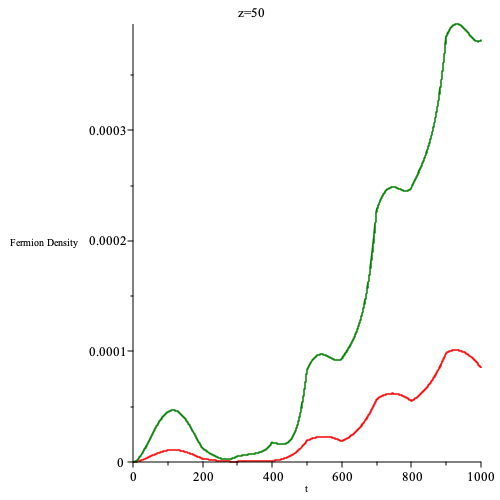}
      \caption{$|\psi_{1L}|^2$ in green and $|\psi_{2L}|^2$ in red without GW as a function of time.}
      \label{fig:fluxwgw}
    \end{subfigure}
    \hfill
    \begin{subfigure}{0.4\textwidth}
      \includegraphics[scale=0.4]{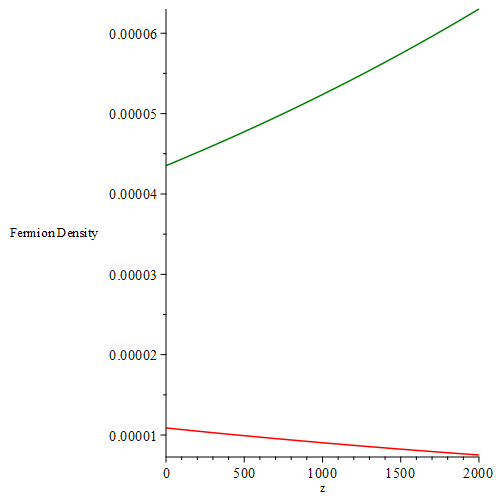}
      \caption{ $|\psi_{1L}|^2$ in green and $|\psi_{2L}|^2$ in red without GW as a function of space at t=600 s.}
      \label{fig:fluxwgw1}
    \end{subfigure}
    \caption{Fermion densities plotted without a GW; We choose $c_1=1, c_2=1/2, g_{ym}= \sqrt{2}, \psi_{1L1}(t,0) = 10^{-2} U(t)= \psi_{2L2}(t,0),$ $ \psi_{1L1}=(0,z)=\psi_{2L2}(0,z)=0$. $\psi_{1L2},\psi_{2L1}$ are set to zero.}
\end{figure}

\begin{figure}[htbp]
     \begin{subfigure}{0.4\textwidth}
      \includegraphics[scale=0.3]{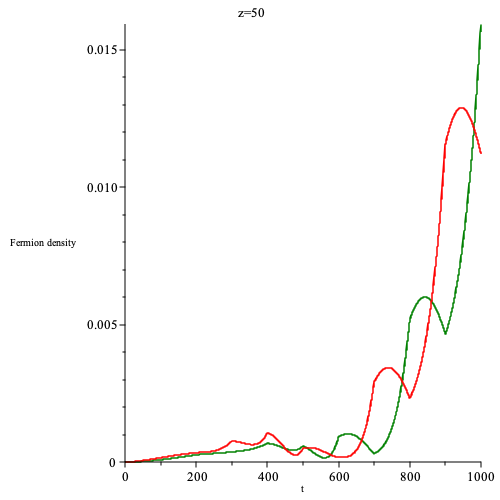}
      \caption{$|\psi_{1L}|^2$ in green and $|\psi_{2L}|^2$ in red with GW as a function of time. }
      \label{fig:fluxwgw12}
    \end{subfigure}
     \hfill
     \begin{subfigure}{0.4\textwidth}
      \includegraphics[scale=0.4]{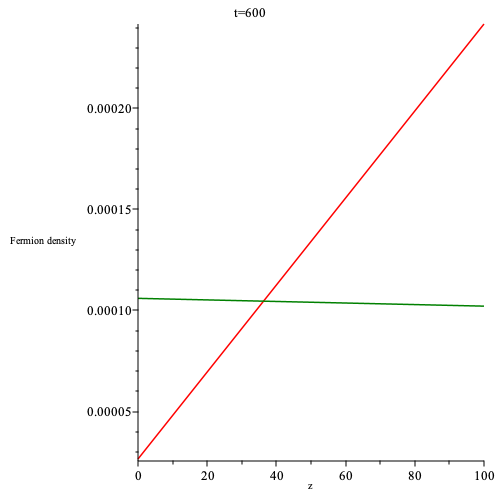}
      \caption{ $|\psi_{1L}|^2$ in green and $|\psi_{2L}|^2$ in redwith GW as a function of space. }
      \label{fig:fluxwgw13}
    \end{subfigure}
    \caption{Figures showing Fermion density with GW as a function of time and space separately. We choose $\omega_g=10^{-4}, A_+= 1, c_1=1, c_2=1/2, g_{ym}= \sqrt{2}, \psi_{1L1}(t,0) = 10^{-2} U(t)= \psi_{2L2}(t,0), $ $ \psi_{1L1}=(0,z)=\psi_{2L2}(0,z)=0$. $\psi_{1L2},\psi_{2L1}$ are set to zero.}
\end{figure}

The $\psi_{2L2}$ component and $\psi_{1L1}$ component of the Fermion $SU(2)$ doublet in the presence of a GW are plotted in Figs. (\ref{fig:psi11} \& \ref{fig:psi21}). What is obvious is that as the GW couples these otherwise independent modes, there should be a flavour transition induced by the GW. 
To analyze the flavour flip behaviour, we have plotted the Fermion density without the GW as a function of $z$ to show that the two flavour densities behave differently, one of them increases but the other decreases in the presence of the condensate as expected from the uncoupled behaviour in Fig. (\ref{fig:fluxwgw1}). With the GW the density shows that there is a coupling effect of the flavours as in Fig. (\ref{fig:fluxwgw13}). Similar behavior is observed for the plots in time as in Figs. (\ref{fig:fluxwgw}) \& (\ref{fig:fluxwgw12}).
In the above, we have taken the amplitude of the GW as 1, which is of the same order as the condensate amplitude. This is to solve for the combined effect of the condensate and the GW without one being a perturbation. This does not imply that the amplitude of the GW is big compared to other fields, as in that case we would have to change Einstein equations. Even though in reality the GW amplitude is usually a perturbation over the matter field, we included this solution to illustrate our results.

If we try to interpret a perturbation due to a weaker GW, then one takes the Fermion wave functions as $\psi_{2L2}+ \tilde{\psi}_{2L2}$ and $\psi_{1L1} + \tilde{\psi}_{1L1}$. The $\tilde{\psi}_{1L1},\tilde{\psi}_{2L2}$ are perturbations proportional to the GW amplitude, $A_+$. The solutions can be obtained using linearized approximations, and the resultant fluxes plotted. It is in this analysis, where the GW is a perturbation over the zeroeth order condensate induced quarks/leptons that we find also the phenomena of `flip'.  We have the following equations, linearized in $h_+$:

\begin{eqnarray}
    i (\partial_0 + \partial_3) \tilde{\psi}_{2L2} -\frac{g_{ym}}{2} U \tilde{\psi}_{2L2}&=& -\frac{g_{ym}}{2} h_+ U \psi_{1L1},
    \label{eqn:pertf1}\\
    i (\partial_0 - \partial_3) \tilde{\psi}_{1L1} - \frac{g_{ym}}{2}U \tilde{\psi}_{1L1} &=& -\frac{g_{ym}}{2} h_+ U \psi_{2L2}. \label{eqn:pertf2}
\end{eqnarray}

\vspace{0.5cm}

\begin{figure}[htbp]
      \includegraphics[scale=0.5]{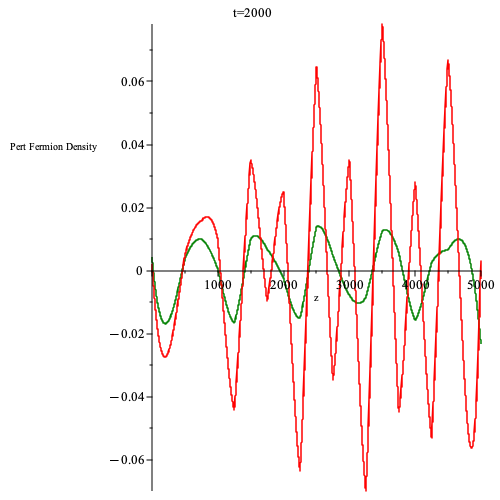}
      \caption{$\frac12(\psi_{1L1}\tilde{\psi}^*_{1L1}+\psi^*_{1L1}\tilde{\psi}_{1L1})$ (green) and $\frac12(\psi_{2L2}\tilde{\psi}^*_{2L2}+\psi^*_{2L2}\tilde{\psi}_{2L2})$ (red) plotted as a function of $z$ with $A_+ = 10^{-2}$ and $\omega_g=1000 $. We choose $c_1=1, c_2=1/2, g_{ym}= \sqrt{2}, \psi_{1L1}(t,0)/2 = 10^{-2} U(t)= \psi_{2L2}(t,0), $ $ \psi_{1L1}=(0,z)=\psi_{2L2}(0,z)=0$. $\psi_{1L2}, \psi_{2L1}$ are set to zero.}
      \label{fig:fluxpertgw}
\end{figure}

\begin{figure}
      \includegraphics[scale=0.4]{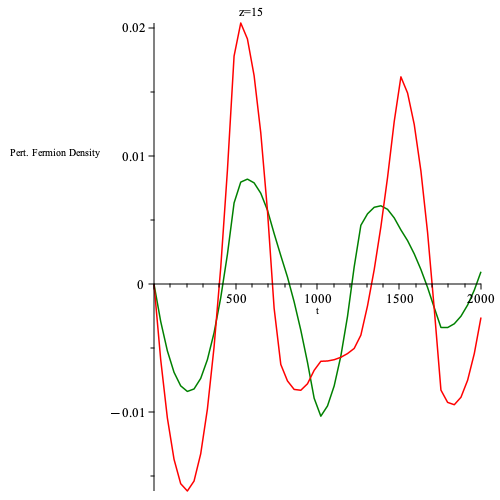}
      \caption{ $\frac12(\psi_{1L1}\tilde{\psi}^*_{1L1} + \psi^*_{1L1}\tilde{\psi}_{1L1})$  (green)
 and $\frac12 (\psi_{2L2}\tilde{\psi}^*_{2L2}+\psi^*_{2L2}\tilde{\psi}_{2L2})$ (red) as a function of time with $A_+ = 10^{-2}$ and $\omega_g=1000 $. We choose $c_1=1, c_2=1/2, g_{ym}= \sqrt{2}, \psi_{1L}(t,0)/2 = 10^{-2} U(t)= \psi_{2L}(t,0), $ $\psi_{1L}=(0,z)=\psi_{2L}(0,z)=0$. The $\psi_{1L2}, \psi_{2L1}$ are set to zero.}
      \label{fig:fluxpertgw21}
\end{figure}

\begin{figure}
      \includegraphics[scale=0.4]{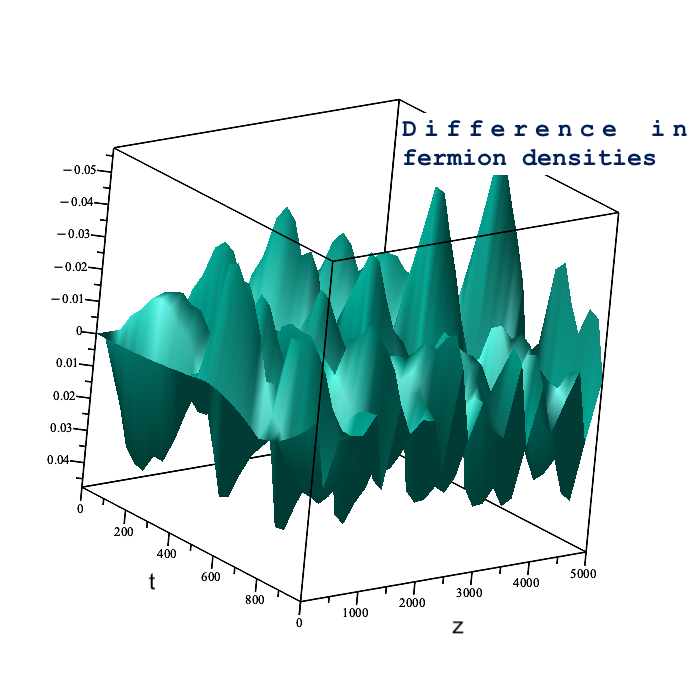}
      \caption{ ($\frac12 (\psi_{1L1}\tilde{\psi}^*_{1L1} +\psi^*_{1L1}\tilde{\psi}_{1L1})-
 \frac12(\psi_{2L2}\tilde{\psi}^*_{2L2} + \psi^*_{2L2}\tilde{\psi}_{2L2})$) as a function of time and space with $A_+ = 10^{-2}$ and $\omega_g=1000 $. We choose $c_1=1, c_2=1/2, g_{ym}= \sqrt{2}, \psi_{1L}(t,0)/2 = 10^{-2} U(t)= \psi_{2L}(t,0), $ $\psi_{1L}=(0,z)=\psi_{2L}(0,z)=0$. The $\psi_{1L2}, \psi_{2L1}$ are set to zero.}
      \label{fig:fluxpertgw1}
    \end{figure}

The solutions to the Eqs. (\ref{eqn:pertf1} \& \ref{eqn:pertf2}) as well as Eqs. (\ref{eq:fermionGW1}, \ref{eq:fermionGW2}) are obtained using numerical methods in MAPLE (worksheet is available on request), and for $g_{ym}=\sqrt{2}$, $c_1=1$ and $c_2=1/2$, the behaviour of the solutions are shown as graphs. What is interesting here is, however, the perturbation density, which has the form $\psi_{1L1}^* \tilde \psi_{1L1} + \tilde{\psi}_{1L1}^* \psi_{1L1}$ and $\psi_{2L2}^* \tilde \psi_{2L2} + \tilde{\psi}_{2L2}^* \psi_{2L2}$, flip at regular intervals both in space and time. We have plotted perturbative fermion density which is equal to $(\psi_{1L1}^* \tilde \psi_{1L1} + \tilde{\psi}_{1L1}^* \psi_{1L1})/2 = {\rm Re}(\psi_{1L1}^* \tilde \psi_{1L1}) $ for $\psi_{1L}$ and similar expression for $\psi_{2L}$. So in other words, if we have $u$ quarks, then interaction with a GW can change the flavour densities, and we would have an enhancement of $d$ quark density. This is because the $u$ quark and the $d$ quark are `coupled' due to the presence of a GW.  This phenomenon of flavour switching is shown in Figs. (\ref{fig:fluxpertgw}, \ref{fig:fluxpertgw21}, \ref{fig:fluxpertgw1}). We also analyze the `difference in the Fermion densities' which flip sign when the GW is introduced in the system.

Note that in the above the Fermion density perturbations are allowed to be negative as they enhance or deplete an already existing zeroeth order flux of the two flavours. We have thus isolated the fluctuations in the flavour density induced by the GW, which is of weaker magnitude than the condensate, and that is the scenario in today's world. If we place the condensate in a $SU(3)$ quark-gluon plasma system, then the GW will induce a transition of the quarks which are coupled by the GW in a $SU(2)$ sub-group. This is an interesting result and can have experimental consequences.  

\subsection{Summary}
In this section, we studied the interaction of $SU(2)$ condensate and Fermion $SU(2)$ doublets. At first, we studied how Fermions react in the presence of condensate.  There is a mixing of some of the Fermion components. We found that the $\psi_{1L1}$ \& $\psi_{2L2}$ (Fermion with flavour 1, Left chirality, component 1; and Fermion with flavour 2, left chirality and component 2) are decoupled and have separate evolution equations (Eqs. (\ref{eqn:fermion1} \& \ref{eqn:fermion4})), but, $\psi_{1L2}$ \& $\psi_{2L1}$ (Fermion with flavour 1, Left chirality, component 2; and Fermion with flavour 2, left chirality and component 1) are coupled to each other described by two coupled partial differential equations (Eqs. (\ref{eqn:fermion2} \& \ref{eqn:fermion3})). When we studied the back-reaction of Fermions on condensate, we found that the back-reaction breaks the isotropy of the YM system. After this, we studied the Fermion + YM condensate system in the presence of a gravitational wave background. We found that there is no change in equations for $\psi_{1L2}$ \& $\psi_{2L1}$, which are still described by Eqs. (\ref{eqn:fermion2} \& \ref{eqn:fermion3}). But, the  $\psi_{1L1}$ \& $\psi_{2L2}$ now become coupled by having a dependence on GW as shown in Eqs. (\ref{eq:fermionGW1} \& \ref{eq:fermionGW2}). We found that the difference in Fermion densities fluctuate in time exhibiting flavour transitions due to GW. It means if we have two Fermions of the same doublet, the Fermion densities of those `flip' at regular intervals. This study of Fermion flip is relevant to QGP and might explain the strangeness of the plasma.
In the above computation, we have shown the enhancement of one flavor in the GW-induced interactions with a quark doublet. This can be implemented for a lepton doublet too. If we focus on a QGP in the presence of a GW, the isospin doublets in our discussion for quarks are $(u,d); (c,s)$, and the above show a flux increase of a $d$ quark, or a $s$ quark in the presence of a GW. In principle the (t,b) pair can also be tested but one would first analyse the interaction with non-zero mass terms to verify the results. In QFT, precisely these interaction vertices can be used to find heavy flavour generation as in \cite{form} from light quarks. The process of generating heavy flavours (strange) using the GW is work in progress \cite{condensate1}.

\section{Conclusion} {\label{sec:5}}
In this article, we have discussed the details of an $SU(2)$ condensate model introduced in \cite{Prokhorov}, in the presence of GW and quarks. We also re-analyzed the results of the energy exchange of the condensate with plasmons. From our re-analysis of the results obtained in \cite{Prokhorov}, we find that their observations that the condensate dissipates into `particles' or `plasmons' is true, but the result is highly dependent on the Yang-Mills interaction constant ($g_{ym}$), and the initial boundary conditions on the plasmons. In our paper, the aim was to study the effect of the GW on the $SU(2)$ condensate. We find that the GW imprints on the condensate generate or induce the plasmons which then decay the condensate. In the gauge invariant ansatz used by \cite{Prokhorov}, we find that the GW interacts only with the symmetric transverse modes identified in their paper. In our re-analysis of the plasmons, the GW interacts with the longitudinal plasmons too. The difference in our results can be attributed to the identification of the correct number of transverse and longitudinal modes of the plasmons using a vector as opposed to a tensor decomposition. We also find that the decay of condensate energy observed in \cite{Prokhorov} is delayed compared to the time observed in \cite{Prokhorov}. The condensate does not decay away all its energy, and the condensate and the plasmons exchange energy in an oscillatory way eventually. This is also shown by the GW+gluon system, which serves to stabilize the plasma. This stabilization of the energy exchange of the plasma-plasmon fields can be used to study the fluid density of the system, and characterize the viscosity of the QGP. It has been shown that the unusually low viscosity of the system can be attributed to the gravitons \cite{grav}, and our model can be used to verify that hypothesis in the presence of GW. 

We then introduce quarks in the background of condensate, and we find that quarks break the isotropy of the condensate. This shows that the quarks + gluon system will have to be described by a non-isotropic field, and this sheds light on the nature of the fluid field required to model the QGP. We also solve the quark-condensate system in the background of a GW and find that the GW can induce a flavour transition of the quarks in case the symmetry group is $SU(2)$. One can also interpret this as the transition of quarks transforming in the same $SU(2)$ sub-group in an $SU(3)$ plasma. We find that without the GW, the quark flavour densities remain decoupled, but the GW couples them. In fact, we find that in the example of the $u$, $d$ flavour doublet, where the $d$ flavour has a lower flux to begin with, the GW causes the $d$ flux to enhance as in Figures (\ref{fig:fluxwgw},\ref{fig:fluxwgw12},\ref{fig:fluxwgw1},\ref{fig:fluxwgw13}). These sort of `flip' in flavour fluxes can be used to generate strangeness observed in the QGP. In \cite{form}, the reactions which give rise to heavy flavour quarks were discussed. Form factors for $q\bar{q}\rightarrow s\bar{s}$ and $gg\rightarrow s\bar{s}$ ($q$ being quark, $g$ being gluons and $s$ being a heavy flavour) were obtained and used in \cite{strange1} to show enhancement of strangeness. In our calculations, the GW-induced flavour transitions represent interaction vertices which have exactly the same structure as the three particle vertices used in \cite{form}, but in a QGP condensate. We therefore expect the contribution of the above interactions to the strangeness generation to be a simple extension. However, we shall be calculating the same process using the GW as an external field instead of a `graviton', and obtaining the enhancement factor \cite{condensate1}.  We are also working on resolving the back-reaction and quantum aspects of the quarks of the system in the work in progress \cite{condensate1}. Our calculations will definitely change the `strangeness' enhancement factors discussed in \cite{strange1} for a QGP formed in the RHIC. In \cite{strange1}, finite temperature versions of the de-confined heavy flavour degeneration is studied. We are currently obtaining the finite temperature distribution functions, and the chemical potentials of the QGP system \cite{condensate1}. When we study the quantization of the above interactions, it would be interesting to probe the novel aspects of entanglement at high energies \cite{ent},\cite{ent1}. Usually, spin conservation is used to study entanglement, but flavour quantum numbers could also be used in the future. Further, most of the work we have presented here discusses the plasmons and Fermions in the presence of the condensate using numerical solutions. Analytic methods such as Floquet linear response theory could be applicable and we will study these techniques next. In summary, the work we have presented here studies the self-interaction of a gluon condensate, quarks, and how GW induces changes in the `plasma' including flavour transitions of the quarks. These classical field theoretic calculations have importance for de-confined YM fields, and they provide the basis for a thermodynamic description of QGP found at high temperatures. Our results show new ways to understand YM fields and GW interactions, we hope that detailed investigations which will provide realistic predictions will be initiated by these.

\vspace{0.5cm}

\noindent
{\bf Acknowledgement:} NRG would like to thank MITACS for the accelerator grant. We would also like to thank the referee for their detailed comments, and for bringing some new suggestions for future work.

\noindent
{\bf Declarations:}
The authors have no competing interests to declare that are relevant to the content of this article. They also declare no conflict of interest with anyone or the funding agencies for conceptualizing and the publication of the results. The authors declare that the results/data/figures have not been published elsewhere, nor are they in consideration for publication by another publisher. The authors declare that the data supporting the findings of this study are available within the paper.

\appendix

\section{Appendix: Method of Lines}{\label{MOL}}
The basic idea of the Method of Lines (MOL) is to replace the spatial derivatives in the PDE with algebraic approximations using finite difference methods. Then, the spatial derivatives will no longer be a function of spatial variables. Now, the only independent variable that remains in all the functions is time. The PDE is approximated by the system of ODEs in time or a system of Differential Algebraic Equations (DAEs). Now, the problem of solving PDE is reduced to the Initial Value Problem (IVP) of the system of ODEs or DAEs \cite{mol}.

Consider a PDE 
\begin{equation}{\label{apeq:pde}}
    \partial_t \partial_t w(t,x) -\partial_x \partial_x w(t,x) = f(t,x).
\end{equation}
First, we need to replace the spatial derivative with an algebraic approximation. We can use the centered finite difference method for the first derivative as 
\begin{equation}
    \partial_x w \approx \frac{w_{i+1}-w_{i-1}}{2 \Delta x},
\end{equation}
where $i$ is an index designating a position along a grid in $x$ and $\Delta x$ is the spacing in $x$ along the grid. Thus, for the left end value of $x$, $i=1$, and the right end value of $x$,$i=N$ i.e. the grid has N points.   
Similarly, for the second-order derivative, it is approximated as  
\begin{equation}
    \partial_x \partial_x w \approx \frac{w_{i+1}-2w_i+w_{i-1}}{ \Delta x^2} .
\end{equation}

Since the equation is second order in $t$  and $x$, it requires two initial conditions (ICs) and two boundary conditions (BCs). Let's consider the following ICs and BCs. 
\begin{align}
    w(t,x=x_i)=g(t),\hspace{1cm}   w(t,x=x_f) = h(t) , \\ w(t=0,x)= g_0(x),\hspace{1cm} \frac{\partial w(t=0,x)}{\partial t} = h_0(x) ,
\end{align}
Using the first BC, we get the value at the first grid point $i=1$ as  
\begin{equation}{\label{eqapp:1}}
    w_1(t)= w(t,x=x_i)=g(t),
\end{equation}
and therefore, we don't need any ODE for $i=1$.

Substituting the approximating expression for derivatives, the PDE (\ref{apeq:pde}) gets transformed as 
\begin{equation}{\label{eqapp:2}}
    \frac{d^2 w_i}{dt^2} = \frac{w_{i+1}-2w_i+w_{i-1}}{ \Delta x^2} + f(t,x_i),\;  i=2,3,...,N.
\end{equation}
For $i=N$, we can use the second BC to get
\begin{equation}{\label{eqapp:3}}
    w_N= w(t,x=x_f)= h(t)
\end{equation}

Finally, we have a DAE system consisting of Eq. (\ref{eqapp:1}) for $i=1$, Eq. (\ref{eqapp:2}) for $i=2,3,..,N-1$ and Eq. (\ref{eqapp:3}) for $i=N$. Thus, we have $N$ number of equations for $N$ unknowns ($w_i,i=1,2,..,N$).
This DAE system can be solved using DAE or ODE solvers in Mathematica.

\section{Appendix: Derivation of the Action in Vector Modes} {\label{AppA}}
The Yang-Mills Lagrangian Density is
\begin{equation}
    {\cal L}= -\frac14 F^a_{\mu \nu} F^{a\mu \nu}.
\end{equation}
Using the Hamilton gauge ($A_0^a=0$), one gets 

\begin{equation}
 \mathcal{L} =\frac12 \partial_0 A_i^a  \partial_0 A_i^a - \frac12 \partial_i A_j^a \partial_i A_j^a + \frac12 \partial_i A^a_j \partial_j A^a_i + g_{ym} \epsilon^{abc} \partial_i A^a_j A^b_i A^c_j
  - \frac{g_{ym}^2}{4} \left((A_i^aA_i^a)^2- (A^a_i A^a_j) (A^b_i A^b_j)\right).
\end{equation}

Next, we assume the vector decomposition as follows
\begin{equation}
    A_i^a= U(t) \delta_i^a + n_i \Phi^a + \epsilon_{ilm} n_l s_m^{\sigma}\chi^a_\sigma.
\end{equation}
where $\sigma =1,2$. We use the following restrictions to get the appropriate splitting of the modes as explained in the main text,
\begin{align}
    \mathbf{n}\cdot \mathbold{\Phi} &=0, \\
    \mathbf{n}\cdot \mathbold{\chi}_\sigma &=0,\\
    \epsilon_{ilm} n_l s_m^{\sigma} \chi^i_\sigma &=0,\;.
\end{align}

This implies that $\Phi^a$ has 2 degrees of freedom, and $\chi^a_{\sigma}$ has 3 degrees of freedom instead of six. In the below, we use a shorthand notation $\Lambda_i^a= \epsilon_{ilm} n_l s_m^{\sigma} \chi^a_\sigma$. 
Using the above gauge field and constraints, the Lagrangian takes the following form
\begin{multline}
    \mathcal{L} = \frac32 (\partial_0 U)^2 + \frac12 n^2 (\partial_0 \Phi^a) (\partial_0 \Phi^a) + \frac12 (\partial_0 \Lambda^a_i) (\partial_0 \Lambda^a_i) - \frac12 n^2 (\partial_i \Phi^a) (\partial_i \Phi^a) -\frac12 (\partial_i \Lambda^a_j) (\partial_i \Lambda^a_j) + \frac12 n_i n_j \partial_i \Phi^a \partial_j \Phi^a \\
     + n_i \partial_j \Phi^a \partial_i \Lambda^a_j + \frac12 (\partial_i \Lambda^a_j)(\partial_j \Lambda^a_i) + g_{ym} \epsilon^{abc} \left[ U^2 n_c \partial_b \Phi^a + U n^2 \Phi^c \partial_b \Phi^a + U n_i n_c \Phi^b \partial_i \Phi^a + U n_c \Lambda^b_i \partial_i \Phi^a \right. \\
    \left. + n^2 \Phi^c \Lambda^b_i \partial_i \Phi^a + U^2 \partial_b \Lambda^a_c + U \Lambda^c_j \partial_b \Lambda^a_j + U n_i \Phi^b \partial_i \Lambda^a_c + n_i \Phi^b \Lambda^c_j \partial_i \Lambda^a_j + U \Lambda^b_i \partial_i \Lambda^a_c + \Lambda^b_i \Lambda^c_j \partial_i  \Lambda^a_j \right]\\
    - \frac14 g^2_{ym} \left[ 6U^4 + 2n^2 U^2 (\Phi^a \Phi^a) + 2U^2 (\Lambda^a_i \Lambda^a_i)- 4 U n^2 \Phi^a \Phi^i \Lambda^a_i - 2U^2 \Lambda^i_j \Lambda^j_i  - 4 U \Lambda^i_j \Lambda^a_i \Lambda^a_j + (\Lambda^a_i \Lambda^a_i)^2 \right. \\
     \left. - (\Lambda^a_i \Lambda^a_j)(\Lambda^b_i \Lambda^b_j) + 2 n^2 (\Phi^a \Phi^a) (\Lambda^b_i \Lambda^b_i) -2 n^2 (\Phi^a \Lambda^a_i) (\Phi^b \Lambda^b_i)    \right].
\end{multline}


To find the explicit equations of motion, we assume some form of the vectors consistent with the restrictions mentioned above as follows
\begin{equation}
        n=(0,0,1) ,\  \  s^1=(1, 0,0), \ \ \ s^2=(0,1,0).
\end{equation}
With this choice, we get from the constraints as
\begin{equation}
    \Phi^3=0, \ \ \chi^3_\sigma = 0,\ \ \chi^1_2 = \chi^2_1.
\end{equation}
We also choose $\chi^2_1 = \chi^1_2 = 0$. With the choice of $\mathbf{n}$, we get the fields to be functions of $t$ and $z$. This reduces the above Lagrangian to
\begin{multline}
     \mathcal{L} = \frac32 (\partial_0 U)^2 + \frac12 (\partial_0 \Phi_1)^2 + \frac12 (\partial_0 \Phi_2)^2 + \frac12 [ (\partial_0 \chi_1)^2 - (\partial_3 \chi_1)^2 ] + \frac12 [ (\partial_0 \chi_2)^2 - (\partial_3 \chi_2)^2 ] \\
    + g_{ym} U^2 \partial_z (\chi_1 + \chi_2)  
    -\frac{g_{ym}^2}{4}\left[ 6U^4 + 2 U^2 (\Phi_1^2 + \Phi_2^2)  - 4 U (\chi_1 -\chi_2) \Phi_1 \Phi_2 + 2 \Phi_1^2 \chi_2^2 \right. \\
      \left. + 2 \Phi_2^2 \chi_1^2
    + 2 U^2 (\chi_1+\chi_2)^2 + 2 \chi_1^2 \chi_2^2\right],
\end{multline}
where $\Phi^1 \equiv \Phi_1$, $\Phi^2 \equiv \Phi_2$, $\chi_1^1\equiv \chi_1 $ and  $ \chi_2^2\equiv \chi_2$. We then use the above Lagrangian to find the equations of motion given in Eqs. (\ref{eq:unogw})-(\ref{eq:chi2}).

If we add the gravitational wave metric to the Lagrangian ($\mathcal{L}$), we get the additional terms ($\mathcal{L}_1$) that are proportional to $h^{\mu \nu}$ which is the contravariant gravitational wave fluctuation over the Minkowski metric ($g^{\mu\nu} = \eta^{\mu \nu}-h^{\mu \nu}$). Those terms are given by 
\begin{align}
  \mathcal{L}_1 = \frac14 F^a_{\mu \nu} \left(h^{\mu \lambda} \eta^{\nu \rho} + \eta^{\mu \lambda} h^{\nu \rho}\right) F^a_{\lambda \rho} .
\end{align}
In Hamilton's gauge with $A_0^a=0$ and the gravitational wave in TT-gauge, $h^{11}= h_+ = A_+ \cos\left(\omega_g (t- z)\right)=-h^{22}$, one gets

\begin{multline}
   \mathcal{L}_1 =  -\frac12 h_+ \Big[\big((\partial_0 A_1^a)(\partial_0 A_1^a) -(\partial_0 A_2^a)(\partial_0 A_2^a)\big) + \eta^{ij} \big(\partial_i A_1^a \partial_j A_1^a - \partial_i A_2^a \partial_j A_2^a\big)\Big] \\
    + \frac{g_{ym}^2 h_+}{4}\Big[ (A^a_1 A^a_1 - A^a_2 A^a_2) A^b_i A^b_i - \eta^{ij}(A^a_1 A^a_i)(A^b_1 A^b_j) + \eta^{ij} (A^a_2 A^a_i) (A^a_2 A^a_j)\Big] .
\end{multline}
         
Using the same vector mode decomposition as before, one gets the following additional terms:
\begin{multline}
    \mathcal{L}_1 =  -\frac12 h_+\left[(\partial_0 \chi_1)^2 -(\partial_0 \chi_2)^2\right]
     + \frac{g_{ym}^2}{4} h_+ \Big[ \chi_2^2 \Phi_1^2 - \chi_1^2 \Phi_2^2 \\ + U^2 (\chi_2^2 -\chi_1^2) + U^2 (\Phi_2^2 - \Phi_1^2) + 2 \Phi_1 \Phi_2 (\chi_1+\chi_2) U \Big] 
\end{multline}
Then, the total Lagrangian in the presence of GW will be $\mathcal{L}+\mathcal{L}_1$.

\section{Appendix: Dirac equation in curved spacetime}{\label{App:Dirac}}
To be clear, we first define the different types of indices. $a,b,c,...$ represents Lie algebra indices, $A,B,C,...$ represents flat spacetime indices, $\mu,\nu,...$ represents curved spacetime indices and $\alpha, \beta$ represents spinor indices. To define the spinors in curved spacetime, we need tetrad formalism. A Tetrad is a set of four linearly independent vectors that can be defined at each point in the curved spacetime. The following relations define the tetrads:
\begin{align}
    g_{\mu\nu} = e^A_\mu e^B_\nu \eta_{AB},\hspace{2cm} & \eta_{AB} = e_A^\mu e_B^\nu g_{\mu\nu},\\
    e^A_\mu e_A^\nu  = \delta^\nu_\mu,\hspace{2cm} &
    e^A_\mu e_B^\mu = \delta^A_B.
\end{align}
where $g_{\mu\nu}$ is the curved spacetime metric, $\eta_{AB}$ is the flat spacetime metric, $e_A^\mu$ are the vector fields and $e^A_\mu$ are the co-vector fields. To write the Dirac equation in the general background, we need to define the spacetime-dependent gamma matrices $\bar{\gamma}^\mu(x)$. These $\bar{\gamma}^\mu$ matrices are related to the familiar constant gamma matrices, $\gamma^A$, by the following relation:
\begin{equation}
    \bar{\gamma}^\mu (x) := e_A^\mu (x)\  \gamma^A,
\end{equation}
where the constant gamma matrices satisfy the $\{\gamma^A,\gamma^B\} = 2 \epsilon \eta^{AB} I $ and the spacetime dependent gamma matrices satisfy $\{\bar{\gamma}^\mu(x),\bar{\gamma}^\nu(x)\} = 2 \epsilon g^{\mu\nu} I $, where $\epsilon=\pm 1$ (depends on metric convention).

With these, we write the Dirac equation as follows
\begin{equation}
    i \bar{\gamma}^\mu (\partial_\mu + \Gamma_\mu) \psi_\alpha + g_{ym} \bar{\gamma}^\mu A^a_\mu T^a_{\alpha\beta} \psi_\beta = 0,
\end{equation}
where $\Gamma_\mu$ is the spinor affine connection given by 
\begin{equation}
    \Gamma_\mu = \frac{\epsilon}{4} w_{AB\mu} \gamma^A \gamma^B, 
\end{equation}
where $w_{AB\mu}$ is the spin connection coefficients given in terms of tetrads as 
\begin{equation}
    w_{AB\mu} = \eta_{AC}\ e^C_\nu\ \nabla_\mu\ e_B^\nu.
\end{equation}

In our case, we are dealing with +-polarised GW propagating in $z$-direction which is given by the following metric:
\begin{equation}
    ds^2=-dt^2 + (1+h_+(t,z)) dx^2 + (1-h_+(t,z)) dy^2 + dz^2,
\end{equation}
With our metric choice, we have $\epsilon =-1$ and one can find the tetrads as follows
\begin{equation}
    e^A_\mu = \begin{pmatrix}
        1 & 0& 0& 0\\
        0 & \sqrt{1+h_+} & 0 & 0\\
        0& 0 & \sqrt{1-h_+} & 0\\
        0 & 0 & 0 & 1 
    \end{pmatrix},
    \hspace{2cm}
    e_A^\mu =  \begin{pmatrix}
        1 & 0& 0& 0\\
        0 & \frac{1}{\sqrt{1+h_+}} & 0 & 0\\
        0& 0 & \frac{1}{\sqrt{1-h_+}} & 0\\
        0 & 0 & 0 & 1 
    \end{pmatrix}
\end{equation}
Using linear order approximation in $h_+$, we can find the spin connection coefficients and spinor affine connection. Plugging everything in the Dirac equation, we get 
\begin{equation}
    i \gamma^0 \partial_0 \psi_\alpha + i \gamma^3 \partial_3 \psi_\alpha + g_{ym} U \left[ \gamma^1 (1-\frac12 h_+) T^1_{\alpha\beta} + \gamma^2 (1+\frac12 h_+) T^2_{\alpha\beta} + \gamma^3 T^3_{\alpha\beta} \right] \psi_\beta = 0
\end{equation}

Using the same assumptions as in without GW case and using the gamma matrices in chiral basis, we get expressions for individual components as follows:
\begin{align}
    i (\partial_0 + \partial_3) \psi_{2L2} + \frac{g_{ym}}{2} h_+ U \psi_{1L1} -\frac{g_{ym}}{2} U \psi_{2L2}&=0,\\
i (\partial_0 - \partial_3) \psi_{1L1} + \frac{g_{ym}}{2} h_+ U \psi_{2L2} - \frac{g_{ym}}{2}U \psi_{1L1} &= 0,\\
i (\partial_0 +\partial_3) \psi_{1L2} - g_{ym} U(t) \psi_{2L1} + \frac{g_{ym}}{2} U(t) \psi_{1L2}&=0,\\
    i (\partial_0 - \partial_3) \psi_{2L1} - g_{ym} U(t) \psi_{1L2} + \frac{g_{ym}}{2} U(t) \psi_{2L1} & =0,
\end{align}

\printbibliography

\end{document}